\documentclass{pasa}%

\def\lpco{{L$^\prime_{\rm CO}$}}
\def\lpcounit{{K.km.s$^{-1}$.pc$^{2}$}}

\usepackage{graphicx}
\usepackage{multirow}
\usepackage{enumerate}

\title[The GLEAMing of the First SMBHs]{The GLEAMing of the First Supermassive Black Holes}

\author[Drouart et al.]{Guillaume Drouart$^1$, Nick Seymour$^1$, Tim J. Galvin$^2$, Jose Afonso$^3$, Joseph R. Callingham$^4$, Carlos De Breuck$^5$, Melanie Johnston-Hollitt$^1$, Anna Kapi\'nska$^6$, Matthew D. Lehnert$^7$ and Jo\"el Vernet$^{5}$
\affil{$^1$International Centre for Radio Astronomy Research, Curtin University, 1 Turner Avenue, Bentley, Western Australia 6102, Australia}
\affil{$^2$CSIRO Astronomy and Space Science, PO Box 1130, Bentley WA 6102, Australia}
\affil{$^3$Instituto de Astrof\'{i}sica e Ci\^{e}ncias do Espa\c co, Universidade de Lisboa, OAL, Tapada da Ajuda, PT1349-018 Lisboa, Portugal}
\affil{$^4$ASTRON, Netherlands Institute for Radio Astronomy, Oude Hoogeveensedijk 4, 7991 PD, Dwingeloo, The Netherlands}
\affil{$^5$European Southern Observatory, Karl Schwarzschild Stra\ss e 2, 85748 Garching bei M\"unchen, Germany}
\affil{$^6$National Radio Astronomy Observatory, 1003 Lopezville Road, Socorro, NM 87801, USA}
\affil{$^7$Sorbonne Universit\'{e}, CNRS, UMR 7095, Institut d'Astrophysique de Paris, 98bis bd Arago, 75014 Paris, France}
}%

\jid{PASA}
\doi{10.1017/pas.\the\year.xxx}
\jyear{\the\year}

\usepackage{aas_macros}
\usepackage{hyperref}
\hypersetup{colorlinks,citecolor=blue,linkcolor=blue,urlcolor=blue}
\usepackage[utf8]{inputenc}

\usepackage[printwatermark]{xwatermark}
\usepackage{xcolor}
\begin{document}

\begin{frontmatter}
\maketitle

\begin{abstract}
We present the results of a new selection technique to identify powerful ($L_{\rm 500\,MHz}>10^{27}\,$WHz$^{-1}$) radio galaxies towards the end of the Epoch of Reionisation. Our method  is based on the selection of bright radio sources showing  radio spectral curvature at the lowest frequency ($\sim 100\,$MHz) combined with the traditional faintness in $K-$band for high redshift galaxies. This technique is only possible thanks to the Galactic and Extra-galactic All-sky Murchison wide-field Array (GLEAM) survey which provides us with 20 flux measurements across the $70-230\,$MHz range. For this pilot project, we focus on the GAMA 09 field to demonstrate our technique. We present the results of our follow-up campaign with the Very Large Telescope, Australian Telescope Compact Array and the Atacama Large Millimetre Array (ALMA) to  locate the host galaxy and to determine  its redshift. Of our four  candidate high redshift sources, we find two powerful radio galaxies in the $1<z<3$ range, confirm one at $z=5.55$ and present a very tentative $z=10.15$ candidate. Their near-infrared and radio properties show that we are preferentially selecting  some of the most radio luminous objects, hosted by massive galaxies very similar to powerful radio galaxies at $1<z<5$.   Our new selection and follow-up technique for finding powerful radio galaxies at $z>5.5$ has a   high $25-50\%$ success rate.
\end{abstract}

\end{frontmatter}

\section{INTRODUCTION }
\label{sec:intro}

For almost four decades up to the 1990s, high redshift radio galaxies were the most distant galaxies known \citep{stern_search_1999}. Since the advent of deep optical surveys and the dropout technique \citep[e.g.][]{steidel_lyman-break_1999}, other selection techniques have overtaken radio, finding both the largest numbers and the most distant galaxies. The stiff competition from optical surveys has also slowed down the search for the most distant radio galaxies; while the radio galaxy TN~J0924-2201 was the second most distant galaxy known at the time of its discovery \citep[$z=5.19$][]{van_breugel_radio_1999}, it was quickly overtaken by many optically selected galaxies. It remained the most distant radio-selected galaxy known for two decades until the discovery of TGSS~J1530+1049 at $z=5.72$ \citep{saxena_discovery_2018}. However, this radio galaxy is less luminous than the bulk of the classical powerful radio galaxies \citep{miley_distant_2008}. While TGSS~J1530+1049 may represent a larger population at these redshifts, the more luminous radio galaxies are important  to study in their own right. Decades of research has demonstrated that the most powerful high redshift radio galaxies are hosted by massive galaxies \citep{seymour_massive_2007,de_breuck_spitzer_2010}, with the most massive black holes \citep{nesvadba_gas_2017,drouart_rapidly_2014} in the most massive dark matter over-densities \citep{galametz_mid-infrared_2012,mayo_overdensities_2012}. Furthermore, should a radio galaxy be detected at $z>6.5$ it would be possible to search absorption by neutral hydrogen in the Epoch of Reionization (EoR) against this background source \citep[e.g.][]{carilli_observations_2004,ciardi_simulating_2015}.

It is difficult to find radio-loud active galactic nuclei (AGN) at the highest redshifts because they are both  optically faint and intrinsically rare. While wide-field optical surveys are now finding quasars out to $z=7.5$ \citep{banados_800-million-solar-mass_2018}, the most distant  powerful ($L_{\rm 500\,MHz}>10^{27}\,$WHz$^{-1}$) radio-loud sources do not currently reach above $z\sim 6$ \citep[e.g.][]{mcgreer_discovery_2006,zeimann_discovery_2011,banados_powerful_2018}. Most radio-loud AGN are obscured type 2 sources, hence are much  harder to find than the bright type 1 (unobscured) AGN. Furthermore, the difficulty in finding such sources is possibly related to the sharp decline in the AGN luminosity function at $z>3$ \citep[e.g.][]{best_cosmic_2014}, the reasons for which are not clear.

As powerful radio galaxies at $z>5$ are so rare, the  best way to find them is from the widest  surveys. Here radio surveys still have an advantage over optical surveys, as several of the former have all-sky coverage. The most difficult part of the identification is that these radio surveys contain hundreds of thousands objects, while their redshift determination remains very time consuming, and can only be done on a few tens of optimal candidates. It is therefore crucial to have an  efficient down-selection directly in the radio. The most popular technique is the selection by means of their ultra-steep radio spectra \citep[e.g.][]{blumenthal_spectral_1979,rottgering_samples_1994,de_breuck_sample_2000,de_breuck_sample_2002,de_breuck_search_2004,broderick_new_2007,saxena_search_2018}. This empirical technique relies on the steepening of the observed spectral indices with redshift, which is due to a fixed concave shape and/or an evolution with redshift of the radio spectral shape \citep[e.g.][]{klamer_search_2006,ker_new_2012}. While these ultra-steep spectrum search techniques have been efficient in finding the most distant radio galaxies, they do offer a biased and incomplete view of the full population of radio-loud AGN as the selection function is not well understood \citep[e.g.][]{barthel_is_1989}. Another disadvantage of this technique is that it requires observations at two frequencies in surveys that are well matched in spatial resolution and depth.

In this paper, we present results from a pilot survey of powerful high redshift radio galaxies using the GLEAM survey\footnote{see \url{http://www.mwatelescope.org/science/gleam-survey}} \citep{wayth_gleam:_2015} observed with the Murchison Widefield  Array \citep[MWA,][]{lonsdale_murchison_2009,tingay_murchison_2013}. The main advantage of this survey is that it provides spectral information  over a wide bandwidth at low frequencies.  Hence, this survey allows one to conduct an efficient selection based on the actual shape of the radio SED, rather than having to rely on a single spectral index. We introduce our new selection technique in \S~2, and then our near-infrared (NIR), the Australian Telescope Compact Array (ATCA) and Atacama Large Millimetre Array (ALMA) follow-up observations in \S~3. In \S~4 we present the first results of our pilot survey. We discuss our results in \S~5, and conclude this paper in \S~6. 

\section{Defining a sample to select z>5 radio galaxies}
\label{sec:cand_selec}

The GLEAM survey is an all-sky low-frequency radio survey performed by the MWA, which provides 20 photometric data-points in the $70-230\,$MHz range. The final resolution is in the $2'-6'$ range for an average rms noise of $\sim$30\,mJy/beam at 151\,MHz.  Sources are detected in a broad-band high-resolution image across $170-230\,$MHz and then flux densities are measured in the 20 8\,MHz sub-bands with priorised positions. The extragalactic first data release \citep{hurley-walker_galactic_2017} contains 307,455 radio components above $5\sigma$ with 99.97\% reliability. We refer the reader to that paper for a complete description of the data processing, source finding and catalogue release. 

We made our original selection using the third internal data release, (IDR3, March 2016) available to the MWA consortium at  that time. Hence the data in Fig.~\ref{fig:selec} and Table \ref{tab:criteria} differ a little from the public data of \cite{hurley-walker_galactic_2017}. These differences are small, but we described them in more detail in Appendix~\ref{sec:IR3_vs_EGCv2}.  Below, we described our selection method.

\subsection{Isolated and compact sources} 

We cross-matched the GLEAM catalogue with the NRAO VLA Sky Survey \citep[NVSS,][]{condon_nrao_1998} catalogue using a cross-matching radius of $30''$. The higher resolution of NVSS ($45''$) allows us to retain only sources that are relatively compact, with a single component and isolated (no other NVSS component within search radius). We apply a flux density cut of 0.4\,Jy$<$$F_{\rm 151\,MHz}$$<$1\,Jy in order to (a) remove the brightest sources in the sample (more likely to be at lower redshift and/or extended) and (b) ensure we are only selecting the most powerful radio galaxies at $z>5$ as well as maintaining sufficient signal-to-noise to reliably fit the MWA SED (see section~\ref{sec:mwafit}). 

\subsection{MWA SED fitting}
\label{sec:mwafit}

We perform a second order polynomial fit in log space to the GLEAM data, following the equation:

\begin{equation}
{\rm log}_{10}\,S_{\nu} = \alpha\,{\rm log}_{10} \left(\frac{\nu_{\rm GLEAM}}{\nu_{0}}\right) + \beta\,{\rm log}_{10} \left(\frac{\nu_{\rm GLEAM}}{\nu_{0}}\right)^2 + \gamma,
\end{equation}

where, $S_{\nu}$ is the flux density in each sub-band, $\nu_{\rm GLEAM}$ is the  corresponding central frequency, $\nu_{0}$ the central frequency of the total GLEAM frequency coverage (151\,MHz) and $\alpha,\beta,\gamma$  are the coefficients of the polynomial terms: $\alpha$ for the spectral index and $\beta$ for the curvature term. The re-normalisation with $\nu_{0}$ permits a curvature estimate within the GLEAM frequency range (70-230\,MHz). We note that a fraction of the sources are better represented by a first order polynomial (better $\chi^2$). However, in the particular case of power-law spectra, the curvature ($\beta$) for these sources will be close to zero and are filtered out during the spectral selection (see next sub-section). 
 
\subsection{Spectral selection}
\label{sec:spec_selec}

\begin{figure} 
    \centering
    \includegraphics[width=0.5\textwidth, trim= 0.5cm 0.5cm 1cm 2.2cm,clip]{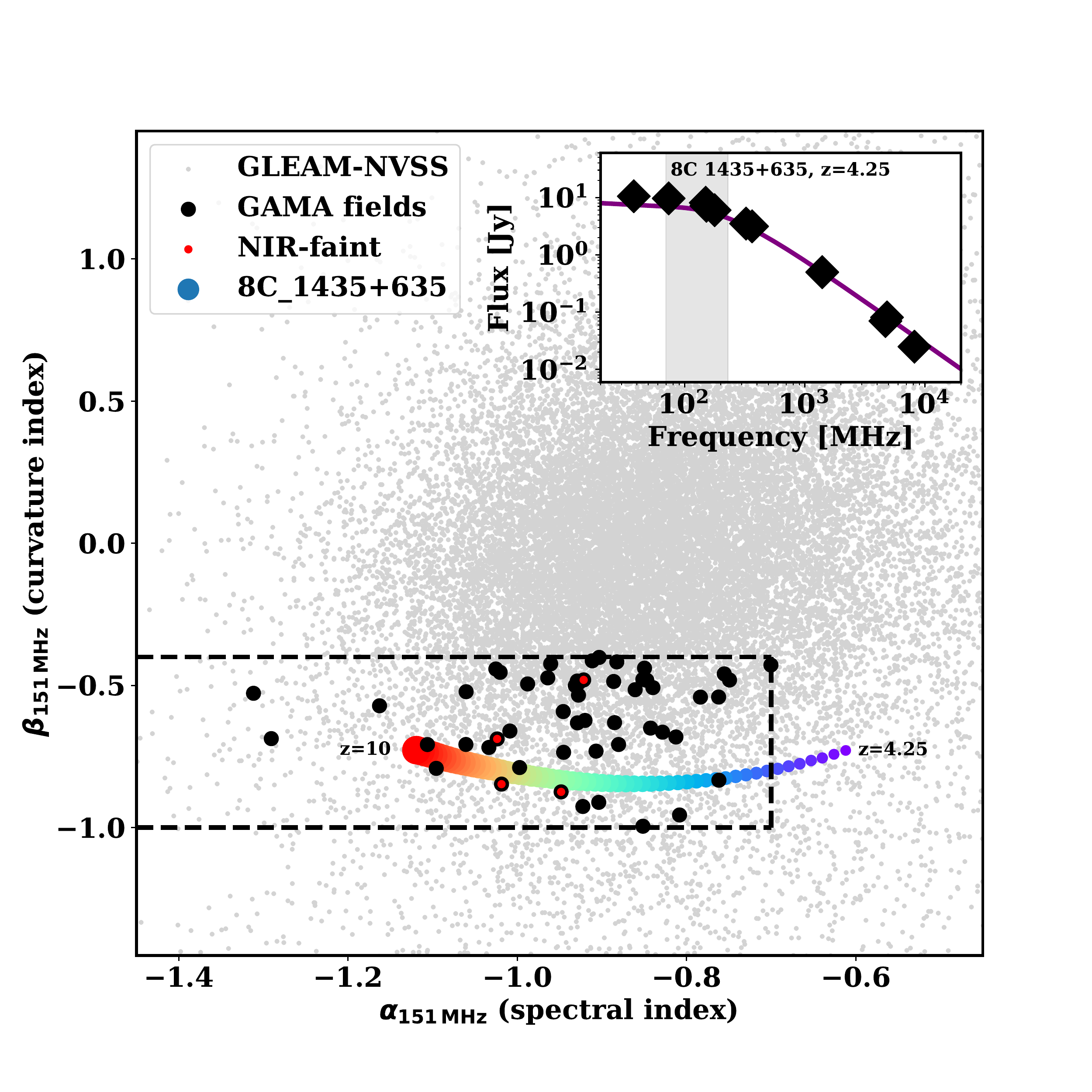}
    \caption{The {GLEAM-derived} spectral index, $\alpha$, versus curvature term, $\beta$, for our full GLEAM-NVSS matched catalogue (grey points). The dashed lines represent our $\alpha$/$\beta$ selection criteria. The black points are the 52 sources in the GAMA survey fields and in red, the four high-redshift candidates with faint or no $K-$band detections  in the GAMA 9\,hour field. The track shows the modelled $\alpha$/$\beta$ values for the powerful radio galaxy 8C~1435+635 (with its SED presented in the inset) when shifted from its redshift of $z=4.25$ up to $z=10$. The grey area in the inset corresponds to the GLEAM frequency coverage.}
    \label{fig:selec}
\end{figure}

In order to further optimise our selection, we make use of the knowledge of one of the largest sample of powerful high redshift radio galaxies to date. A significant fraction of the sources at $z>3.5$ in the ``HeRG\'E'' sample \citep{seymour_massive_2007,de_breuck_spitzer_2010}, present a flattening in their radio SED when moving to the lowest frequencies (Drouart et al. in prep).  We use one of these sources, 8C 1435+635 (at $z=4.25$ see the inset in Fig.~\ref{fig:selec}) to predict the $\alpha$ and $\beta$ at higher redshift (coloured line in Fig.~\ref{fig:selec}). We see that the track passes through the $-1.0<\beta<-0.4$ region, and generally presents a steep spectral index with $\alpha<-0.7$ (expected as the sources in the HeRG\'E sample were selected for their steep spectrum and radio luminosity). We therefore isolate this part of the parameter space to select our candidates. We also apply the  criteria of $\chi^2<5$ and $|\beta/\delta\beta|>1$, to ensure that we select source with robust fitting and a significantly curved SED.

\begin{table}
    \caption{Summary of our selection criteria conducted on the IDR3 GLEAM catalogue. Using the public GLEAM release will give slightly different numbers (see Appendix~\ref{sec:IR3_vs_EGCv2}). See \S~\ref{sec:spec_selec} for more details.}
    \centering
    \begin{tabular}{lr}
    \hline 
        Criteria & \# sources \\
         \hline \hline
        GLEAM IDR3 & 304\,894 \\
        NVSS counterparts at $<30''$ &  217\,877 \\
        Single NVSS $+$ compact ($\theta_{min}<20''$) & 127\,846 \\
        0.4\,Jy$<F_{\rm 151}<$1.0\,Jy &  27\,597 \\
        $\chi^2$$<$5 $+$ significant curvature ($|\beta/\delta\beta|$$>$1) & 10\,536 \\
        $\alpha/\beta$ selection (see \S~\ref{sec:spec_selec} and Fig.~\ref{fig:selec}) &  2\,338 \\
        in GAMA fields & 52 \\
        in GAMA~09 field $+$ no $K-$band  & 4 \\
        \hline
    \end{tabular}
    \label{tab:criteria}
\end{table}

\begin{table*}
    \caption{Summary of our candidate high redshift radio galaxies including original IDR3 name, final GLEAM name (both preceded by `GLEAM'), short name for this paper, IDR3 151\,MHz flux density and the best fit $\alpha/\beta/\chi^2$ parameters using the IDR3 data. Note the source positions are given by their name with uncertainties of $3-5''$.}
    \centering
    \begin{tabular}{llc c ccc}
    \hline 
    IDR3  name &  final  name & Short name  & $F_{\rm 151MHz}$ [mJy] & $\alpha$ & $\beta$ & $\chi^2$ \\
        % we need the final GLEAM name which differs from the IDR3 name
    \hline \hline
    J085614$+$022400 & J085614$+$022359 & GLEAM 0856  & $905\pm30$ & $-1.02\pm0.04$ & $-0.84\pm0.26$ & 0.23 \\
    J091337$+$023154 & J091337$+$023154 & GLEAM 0913 & $520\pm30$ & $-0.92\pm0.06$ & $-0.48\pm0.38$ & 0.36 \\
    J091734$-$001243 & J091734$-$001243 & GLEAM 0917  & $473\pm28$ & $-0.94\pm0.08$ & $-0.87\pm0.59$ & 0.67 \\
    J091823$-$000509 & J091823$-$000509 & GLEAM 0918  & $441\pm28$ & $-1.02\pm0.07$ & $-0.68\pm0.49$ & 0.53 \\
    \hline
    \end{tabular}
    \label{tab:sample}
\end{table*}

\subsection{Pilot project on the GAMA 09 field}

Finally, we focus on the  Galaxy And Mass Assembly (GAMA) survey  \citep{driver_gama:_2009} where a wealth of multi-wavelength data is available to further isolate the most promising candidates. When considering the five GAMA fields together, only 52 sources are left. 
We visually inspected each source with a $JHK$ image from the VIKING survey \citep[][]{arnaboldi_eso_2007}. The VIKING 5$\sigma$ sensitivity, mag $K_s(\rm AB)=22.1$, is ideally suited to eliminate any lower redshift sources due to the well established $K-z$ relationship \citep{rocca-volmerange_radio_2004} for
powerful radio galaxies. 
After isolating sources which present no, or very faint NIR, counterparts within a few arc-seconds of the GLEAM position,  and restricting ourselves to the GAMA 9 hour field (hereafter G09) in order to optimise our ALMA follow-up, we are left with four high-$z$ candidates  (see Table~\ref{tab:sample}). 

To identify the host galaxies, we obtained the following observations: (i) a moderately deep $K_s-$band imaging with the HAWK-I instrument on the Very Large Telescope (VLT) in order to confirm the location of the host in NIR, not detected in the VIKING data (see \S~\ref{sec:HAWK-I}), (ii) ATCA continuum imaging at 5.5, 9.0, 17 and 19\,GHz to complete the radio SED coverage to over four decades in frequency (see \S~\ref{sec:data_atca}) and, (iii) an ALMA Band 3 spectral scan (84-115\,GHz, \S~\ref{sec:data_alma}) to obtain simultaneously a deep, sub-arcsecond resolution continuum image to identify the host galaxy in the HAWK-I $K_s$ imaging, and to obtain the redshift via the molecular emission lines --- the method the South Pole Telescope (SPT) team used for strongly lensed sub-millimetre (sub-mm) galaxies \citep{weis_alma_2013,strandet_redshift_2016}.

\section{Follow-up Data}

\begin{figure*}
    \centering
    \begin{tabular}{cc}
        \includegraphics[width=0.48\textwidth,trim=0 0 0 0,clip]{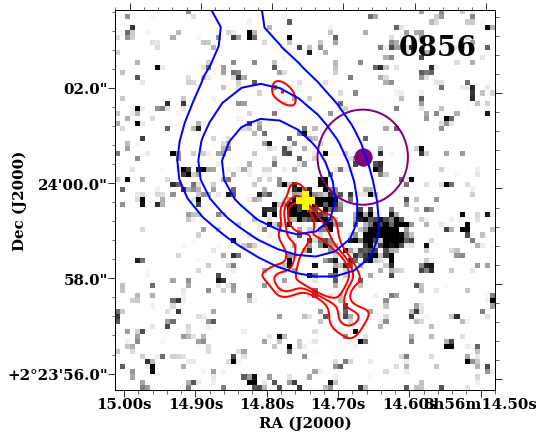} & \includegraphics[width=0.48\textwidth,trim=0 0 0 0,clip]{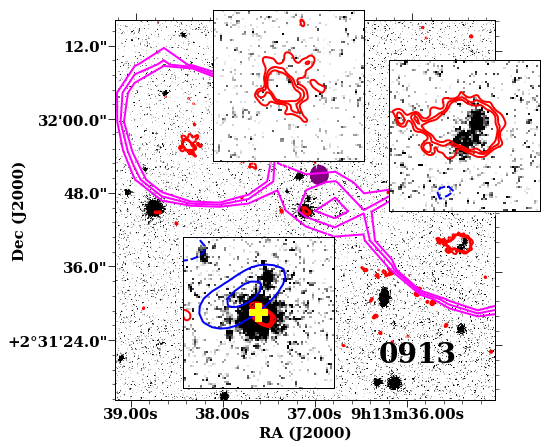} \\
        \includegraphics[width=0.48\textwidth,trim=0 0 0 0,clip]{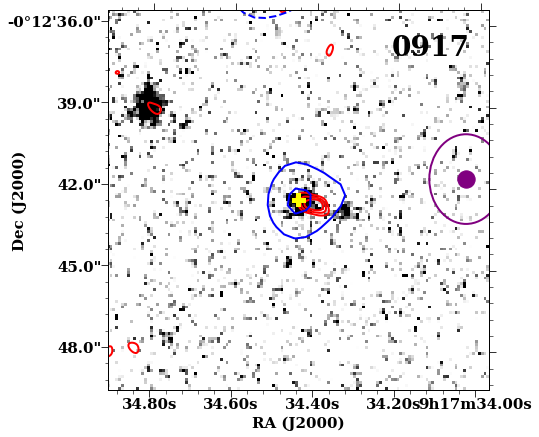} & \includegraphics[width=0.48\textwidth,trim=0 0 0 0,clip]{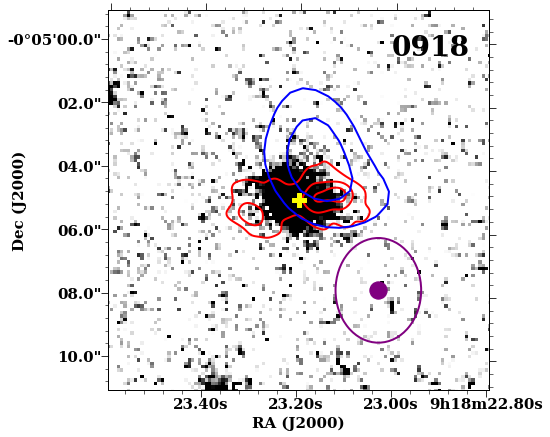} \\
    \end{tabular}
    \caption{$K_s-$band gray-scale VLT/HAWK-I images of our four GLEAM-selected targets with {\bf the continuum and line ALMA data overlaid as red and blue contours, respectively}. The yellow cross indicates the coordinates for the ALMA spectra presented in Fig.~\ref{fig:ALMA_spec}, and the purple circles are the GLEAM IDR3 position with their uncertainties as ellipses. The red contours represent the ALMA continuum emission (at 3, 4, 5, 10, 15$\sigma$ with $\sigma$=9$\,\mu$Jy/beam).   later). 
    The blue contours show, at lower resolution (see \S~\ref{sec:data_alma} and \S~\ref{sec:morph}), the integrated detected lines in the spectra (see Fig.~\ref{fig:ALMA_spec}) as follows (solid for positive signal, dashed for negative). 0856 ({\it top left}), the blue contours are the average-stacked CO emission at 2$\sigma$, 3$\sigma$ and 4$\sigma$ with $\sigma$=140$\,\mu$Jy/beam. 0913 ({\it top right}), the blue contours are the detected line at 108 (at -2, 2, 3$\sigma$ with $\sigma$=140$\,\mu$Jy/beam, see also \S~\ref{sec:z_deter}). Note this image covers a much larger field of view ($\sim 1'$) compared to other images. We overlay the 19\,GHz ATCA image as magenta contours at 3, 4, 5$\sigma$ with $\sigma$=70$\,\mu$Jy/beam. The insets show a close-up of the core and lobes. 0917 ({\it bottom left}), the blue contours are the average-stacked CO lines at 2, 3$\sigma$ with $\sigma$=160$\,\mu$Jy/beam. 0918 ({\it bottom right}), the blue contours shows the CO line at 2, 3$\sigma$ with $\sigma$=160$\,\mu$Jy/beam.}
    \label{fig:morph}
\end{figure*}

\begin{table*}
\begin{center}
\caption{NIR flux densities from our HAWK-I $K_s$ and the VIKING $zYJHK$ images.  The aperture photometry radius is defined from the HAWK-I $K_s-$band images. The upper limits are the 3$\sigma$ values from the corresponding image. Flux densities are not corrected for galactic extinction which is very low towards the GAMA 09 field.}
\begin{footnotesize}
\begin{tabular}{lllccccccc}
\hline
Name & RA & Dec & Ap. rad. & $z$ & $Y$ & $J$ & $H$ & $K$ & HAWK-I-$K_s$ \\
& [J2000] & [J2000] & [arcsec] & [$\mu$Jy] & [$\mu$Jy] & [$\mu$Jy] & [$\mu$Jy] & [$\mu$Jy] & [$\mu$Jy] \\
\hline \hline
0856  & 8:56:14.73 & 2:23:59.6 & 0.8 & 0.52$\pm$0.16 & $<$1.12 & $<$2.11 & $<$4.11 & $<$2.85 & 1.88$\pm$0.11 \\ 
0913  & 9:13:37.14 & 2:31:45.4 & 2.0 & 7.51$\pm$0.51 & 11.25$\pm$1.19 & 17.47$\pm$2.22 & 21.32$\pm$1.88 & 43.12$\pm$2.31 & 37.81$\pm$0.27 \\ 
0917  & 09:17:34.36 & -00:12:42.7 & 1.0 & $<$0.24 & $<$0.53 & $<$0.63 & $<$1.26 & $<$3.87 & 3.07$\pm$0.12 \\
0918 & 9:18:23.2 & -0:05:05 & 3.0 &  4.34$\pm$0.57 & 4.97$\pm$1.26 & 10.79$\pm$1.86 & 17.73$\pm$3.94 & 32.16$\pm$3.67 & 25.51$\pm$0.44 \\ 
\hline
\end{tabular}
\end{footnotesize}
\label{tab:phot}
\end{center}
\end{table*}

 \begin{figure*}
    \centering
    \begin{tabular}{cc}
        \includegraphics[width=0.49\textwidth,trim=0 0 0 0,clip]{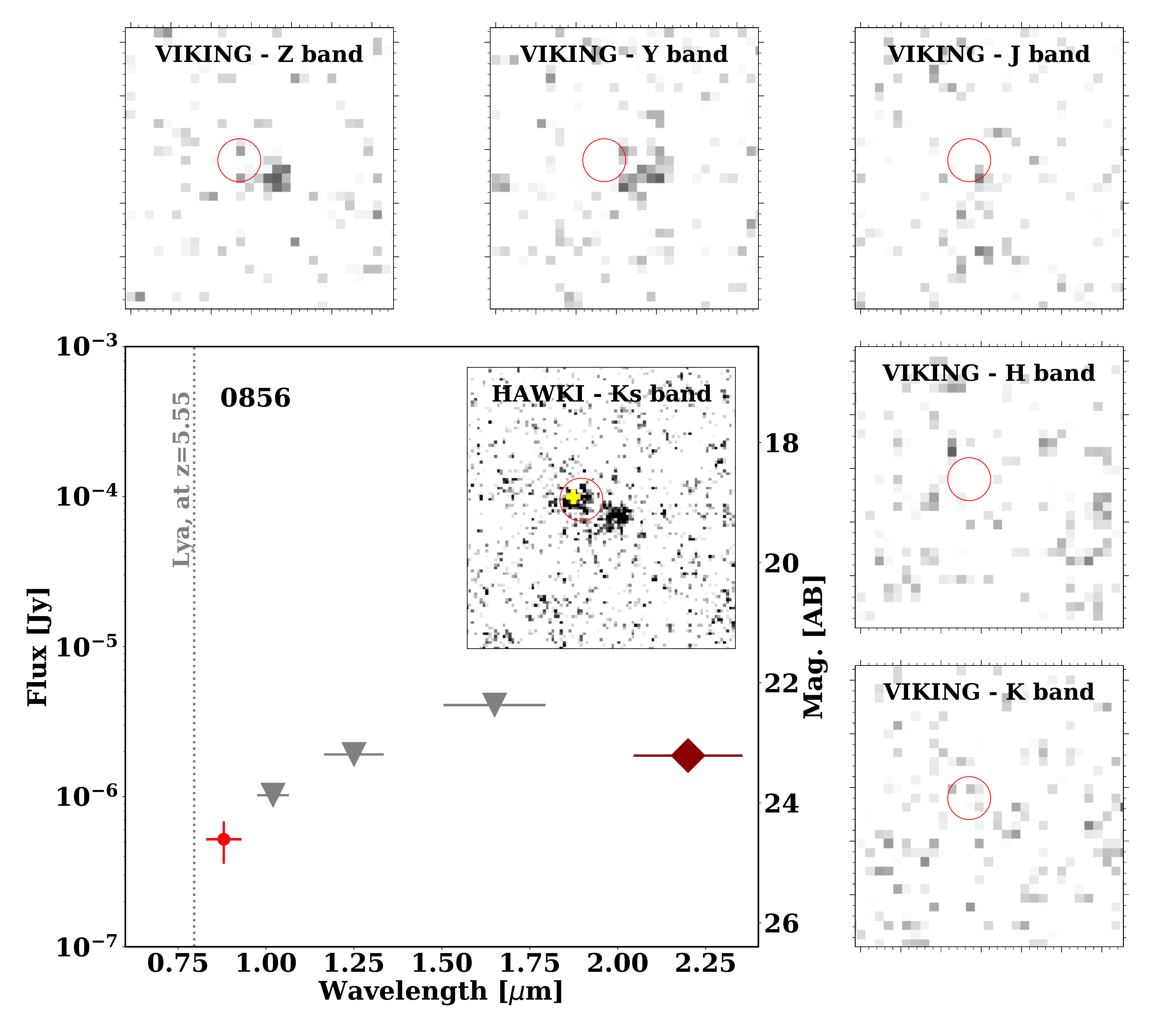} & \includegraphics[width=0.49\textwidth,trim=0 0 0 0,clip]{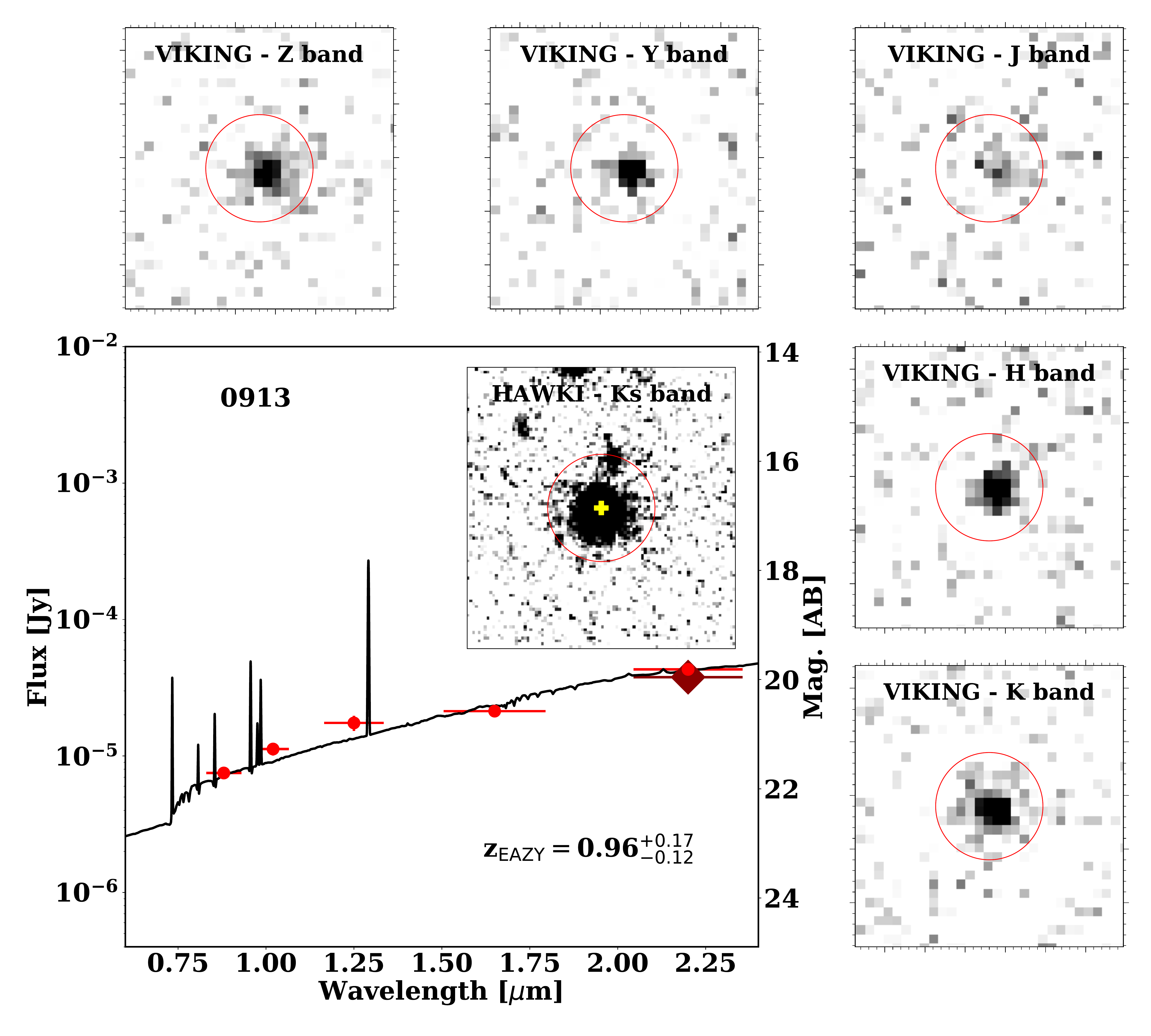} \\
        \includegraphics[width=0.49\textwidth,trim=0 0 0 0,clip]{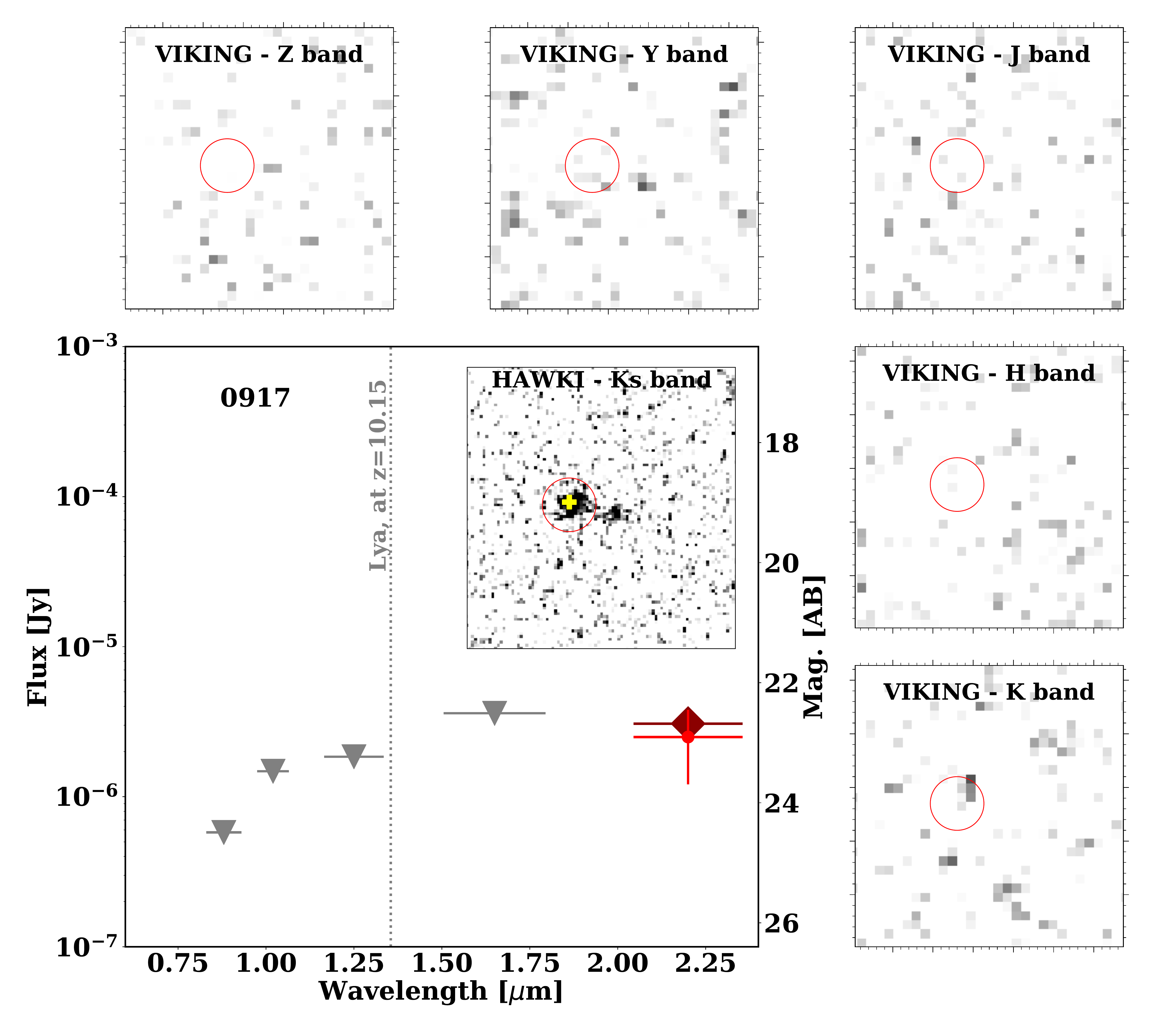} & \includegraphics[width=0.49\textwidth,trim=0 0 0 0,clip]{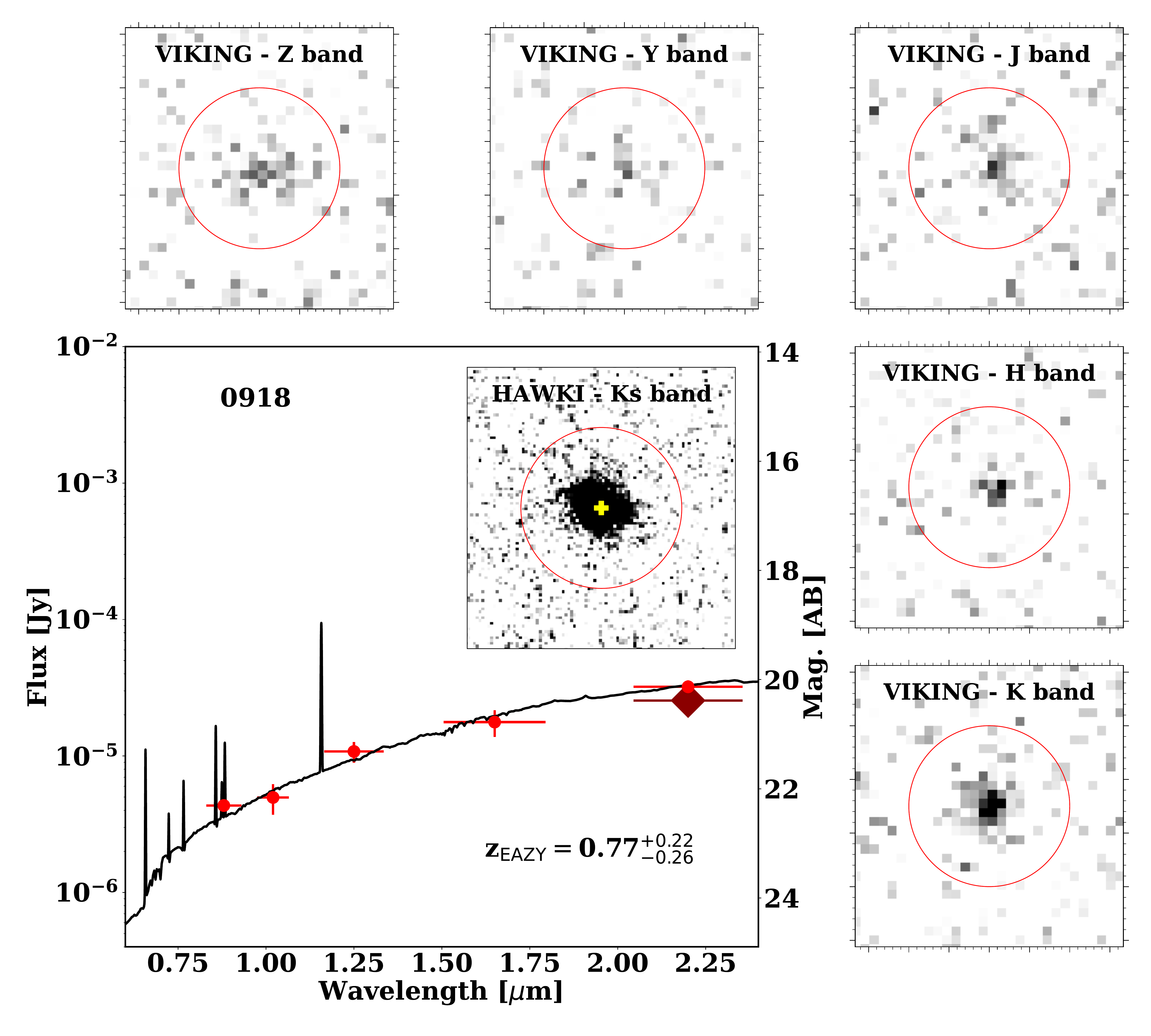} \\
    \end{tabular}
    \caption{The NIR SEDs for our four sources with insets presenting  each of the corresponding NIR images. We report the flux density (or $3\,\sigma$ upper limit) for each band in Table~\ref{tab:phot}. The circles in the insets are the apertures defined from the HAWK-I images (close insert) and also applied to the VIKING images (outer inserts). The dark large diamonds are the HAWK-I measurements. The grey symbols report the VIKING 3$\sigma$-sensitivity in the case of non-detections. The horizontal error bars are the FWHM of the bands ($zYJHK$). The yellow crosses indicate the coordinates for the extraction of the ALMA spectra presented in Fig.~\ref{fig:ALMA_spec}. For the two sources with potential confirmed redshifts (0856 and 0917) we indicate the observed wavelength of the Lyman$-\alpha$ line. For the two other sources (0913 and 0918) we present the best fit model for photometric redshift determination performed with EAZY (see \S~\ref{sec:photz}). }
    \label{fig:sed_gama}
\end{figure*}

 \subsection{NIR follow-up observations}
\label{sec:HAWK-I}

Our ESO/HAWK-I programme (0101.A-0571A, PI Drouart) was observed between 2018 April and June in service mode. It consisted of two sets of exposure in $K_s$-band, centred on CHIP1 (bottom-left quadrant). The total exposure time on each source reaches 1\,h. We processed the data with the {\tt esorex} tool and followed the standard processing recipe described in the HAWK-I manual v2.4.3 for each run. The final mosaics were created using the {\tt Montage} package v5.0.0 and {\tt montage-wrapper} v0.9.9 with all options set to their default values. We present the $K_s-$band images of our sources in Fig.~\ref{fig:morph}. Finally, we perform an aperture photometry based on the 
 positions of host galaxies found in the HAWK-I images and confirmed by the ALMA continuum emission.
Flux densities at these positions were extracted from the VIKING $zYJHK$ bands using the {\tt phot.utils} package v0.5. These flux densities are reported in Table \ref{tab:phot} and the resultant SEDs presented in Fig.~\ref{fig:sed_gama}.
a
\begin{table*}[t]
\caption{Radio flux density summary for our four radio galaxies. The reported GLEAM values below are from the public release, not IDR3. $^s$the flux densities are the sum of multiple components.}
\centering
\label{tab:radio}
\begin{tabular}{ccccccc}
\hline
Name & beam & Frequency  & Flux 0856 & Flux 0913 & Flux 0917 & Flux 0918 \\
     & [arcsec] & [GHz] & [mJy] & [mJy] & [mJy] & [mJy] \\
\hline \hline
GLEAM & 316.0 & 0.076 & 1400 $\pm$ 81.4 & 950 $\pm$ 85.8 & 673 $\pm$ 92.2 & 839 $\pm$ 92.4 \\ 
GLEAM & 286.0 & 0.084 & 1320 $\pm$ 64.9 & 912 $\pm$ 66.7 & 672 $\pm$ 69 & 663 $\pm$ 73.9 \\ 
GLEAM & 265.0 & 0.092 & 1250 $\pm$ 55.4 & 738 $\pm$ 59.7 & 711 $\pm$ 61.2 & 574 $\pm$ 60.8 \\ 
GLEAM & 247.0 & 0.099 & 1170 $\pm$ 49.6 & 672 $\pm$ 54 & 714 $\pm$ 53.6 & 552 $\pm$ 54.4 \\ 
GLEAM & 230.0 & 0.107 & 1200 $\pm$ 57.1 & 743 $\pm$ 54.5 & 541 $\pm$ 51 & 540 $\pm$ 51.3 \\ 
GLEAM & 213.0 & 0.115 & 1060 $\pm$ 45.5 & 605 $\pm$ 47.3 & 589 $\pm$ 44.2 & 564 $\pm$ 44.4 \\ 
GLEAM & 201.0 & 0.122 & 1000 $\pm$ 42.3 & 660 $\pm$ 43.6 & 612 $\pm$ 38.1 & 524 $\pm$ 38.4 \\ 
GLEAM & 190.0 & 0.13 & 903 $\pm$ 40.1 & 613 $\pm$ 41.7 & 561 $\pm$ 37.7 & 508 $\pm$ 37 \\ 
GLEAM & 175.0 & 0.143 & 871 $\pm$ 31.6 & 596 $\pm$ 31.4 & 563 $\pm$ 30.9 & 465 $\pm$ 30.7 \\ 
GLEAM & 166.0 & 0.151 & 894 $\pm$ 27.1 & 520 $\pm$ 29.3 & 465 $\pm$ 27 & 437 $\pm$ 27.4 \\ 
GLEAM & 159.0 & 0.158 & 778 $\pm$ 26.5 & 547 $\pm$ 25.8 & 442 $\pm$ 25.3 & 421 $\pm$ 25.9 \\ 
GLEAM & 152.0 & 0.166 & 816 $\pm$ 25 & 502 $\pm$ 28.5 & 434 $\pm$ 25.9 & 372 $\pm$ 24.9 \\ 
GLEAM & 147.0 & 0.174 & 732 $\pm$ 27.9 & 493 $\pm$ 28.9 & 390 $\pm$ 24.7 & 384 $\pm$ 24.8 \\ 
GLEAM & 140.0 & 0.181 & 705 $\pm$ 23.8 & 479 $\pm$ 24.8 & 322 $\pm$ 22 & 374 $\pm$ 22.3 \\ 
GLEAM & 135.0 & 0.189 & 685 $\pm$ 26.7 & 431 $\pm$ 26 & 417 $\pm$ 23 & 340 $\pm$ 22.2 \\ 
GLEAM & 130.0 & 0.197 & 583 $\pm$ 26.4 & 422 $\pm$ 26.4 & 369 $\pm$ 22.5 & 282 $\pm$ 22.6 \\ 
GLEAM & 124.0 & 0.204 & 585 $\pm$ 23.4 & 380 $\pm$ 27.3 & 353 $\pm$ 24.2 & 249 $\pm$ 23.5 \\ 
GLEAM & 121.0 & 0.212 & 583 $\pm$ 21.7 & 415 $\pm$ 26.2 & 276 $\pm$ 21.4 & 252 $\pm$ 21.5 \\ 
GLEAM & 117.0 & 0.22 & 521 $\pm$ 20.1 & 326 $\pm$ 21 & 298 $\pm$ 20.1 & 277 $\pm$ 20.1 \\ 
GLEAM & 114.0 & 0.227 & 532 $\pm$ 19.5 & 335 $\pm$ 21.1 & 318 $\pm$ 19.7 & 279 $\pm$ 18.3 \\ 
TGSS & 26.4 & 0.15 & 870 $\pm$ 87 & 549 $\pm$ 55.4 & 486 $\pm$ 48.8 & 463 $\pm$ 46.5 \\ 
NVSS & 17.6 & 1.4 & 86.5 $\pm$ 2.6 & 94.3 $\pm$ 3.2 & 46.6 $\pm$ 1.5 & 54.8 $\pm$ 1.7 \\ 
ATCA & 41.2 & 5.5 & 15.5 $\pm$ 1.55 & 31 $\pm$ 3.1 & 7.68 $\pm$ 0.768 & 14.2 $\pm$ 1.42 \\ 
ATCA & 25.0 & 9 & 7.64 $\pm$ 0.764 & 20.2 $\pm$ 2.02 & 3.53 $\pm$ 0.353 & 8.09 $\pm$ 0.809 \\ 
ATCA & 14.0 & 17 & 2.92 $\pm$ 0.292 & 9.45 $\pm$ 0.945 & 1.22 $\pm$ 0.122 & 3.74 $\pm$ 0.374 \\ 
ATCA & 11.5 & 19.4 & 2.08 $\pm$ 0.21 & 8.5 $\pm$ 0.85 & 0.93 $\pm$ 0.12 & 3.18 $\pm$ 0.32 \\ 
ALMA & 0.8 & 100 & 0.261 $\pm$ 0.0216 & 1.78 $\pm$ 0.033 & 0.0788 $\pm$ 0.0086 & 1.38 $\pm$ 0.04 \\
\hline
\end{tabular}
\end{table*}

 \subsection{ATCA follow-up observations}
\label{sec:data_atca}

 To fill the spectral gap between 1.4\,GHz and the ALMA observations at 100\,GHz, we followed up the sources with the ATCA \citep[][]{frater_australia_1992} over a five hour period on the 02-Dec-2018 under the project code CX420. Using the Compact Array Broadband Backend \citep[CABB][]{wilson_australia_2011}, 2.048\,GHz bandwidth observations were conducted at nominal central frequencies of 5.5, 9.0, 17.0 and 19.0\,GHz in the compact H168 array configuration. For the 5.5 and 9.0\,GHz observations, we used PKS\,1934-638 to establish an absolute flux density consistent with the \cite{baars_reprint_1977} standard \citep{partridge_absolute_2016} as well as to derive our bandpass correction. For the  17.0 and 19.0\,GHz data, we used PKS\,B0420-014 to produce our bandpass correction. Across all observing frequencies, we targeted 0906+015 for phase calibration at least once every ten minutes. 

The data were calibrated and imaged using the \textsc{miriad} software package \citep{sault_retrospective_1995} following standard procedures. 
They were loaded into \textsc{miriad} using \textsc{atlod} and  flagged for radio RFI using the guided automated flagging \textsc{pgflag} task. During this stage it was revealed that much of the 19.0\,GHz data suffered from a broadband RFI, likely from a nearby satellite. As a consequence about 1,000 channels were flagged near the centre of the band. Next an initial bandpass and flux scale were established for the 5.5 and 9.0\,GHz data with \textsc{mfcal} and \textsc{gpcal} using the calibrator PKS\,1934-638 as a reference spectrum. These solutions were then \textsc{gpcopy}ed over to the 0906+015 and time varying phase solution was estimated using \textsc{gpcal}. Finally, \textsc{mfboot} was used to correct for any percentage level flux scaling mismatch, and these solutions were used for the four sources. A similar procedure was used when calibrating the 17.0 and 19\,GHz data, except \textsc{mfcal} used PKS\,B0420-014 to produce bandpass corrections. These were subsequently copied over to PKS\,1934-638 and steps following the procedure outlined above were carried out. To allow any frequency dependent terms to be accounted for, \textsc{nfbin} was set to two for all appropriate calibration tasks.

Images were created using \textsc{invert} with a Briggs \textsc{robust} parameter of 2 (equivalent to a natural weighting). We did not include the 6$^{\mathrm{th}}$ ATCA antenna, separated by $\sim4.5$\,km from the central core of the H168 array. Owing to the large fractional bandwidth of the data, the wideband imaging deconvolution task \textsc{mfclean} \citep{sault_multi-frequency_1994} was used. Next, \textsc{restor} and \textsc{linmos} were used together to deconvolve telescope artefacts and apply primary beam corrections while accounting for the spectral index terms constrained by \textsc{mfclean} for each of the clean components. The final sensitivity on the images reaches typically the 20-60$\,\mu$Jy/beam level and the final synthesised beam goes from $48''\times 34''$ to $13''\times 9''$ from 5.5 to 19\,GHz, respectively. Finally, we ran Aegean \citep{hancock_source_2018} on each image to extract the flux observed in the continuum images and report the total flux densities in Table~\ref{tab:radio}.

\subsection{ALMA Band 3 follow-up observations}
\label{sec:data_alma}

As our four sources were close together in G09 we could optimise our observing strategy by sharing overheads. Our main goal is to detect one or more CO lines in order to determine a secure redshift. At $z>5$ we expect two or more lines in the full frequency range. Our project 2017.1.00719.S was observed on 2018 January 25 and 26 in configuration C43-5 (total $\sim$6\,h) and a with a high perceptible water vapour of $PWV$=5.2\,mm and $PWV$=6.5\,mm respectively. We requested a spectral scan in Band 3, covering the 84-115\,GHz range with five different tunings. All processing is performed with CASA in version 5.3.0 \citep{mcmullin_casa_2007}. We inspect the visibilities to check for extra-flagging with the {\tt plotms} CASA routine, merge the five sub-cubes at different frequencies (tunings) into a single cube per source. 

\subsubsection{Continuum image and flux extraction}

We created a {\bf 30\,GHz-continuum (total ALMA bandwidth) centered at $\nu_{\rm sky}=$100\,GHz image} using natural weighting in order to maximise sensitivity for each source. The final images present an averaged synthesised beam of $0.7''\times0.8''$, a noise of about 9$\,\mu$Jy/beam and are presented as red contours in Fig.~\ref{fig:morph}. We ran Aegean with all defaults parameters to derive the photometry and report the results in Table~\ref{tab:radio}.

\subsubsection{Spectral cubes, spectrum extraction and line detection}

As the expected line strength was low and the atmospheric conditions non-optimal (high PWV, see \S~\ref{sec:data_alma}), we decided to taper our data by a $3''$ Gaussian beam in order to minimise the noise contribution of the longest baselines (our longest baselines are 1.4\,km) in order to increase our signal to noise for detection. We also smoothed the data in frequency space from 20 to 80\,MHz channels. The corresponding final sensitivity in the inverted cubes is $\sim0.3\,$mJy/beam for 80\,MHz-width channels. 

\begin{figure*}
    \centering
    \includegraphics[width=1.0\textwidth,trim=100 80 100 100,clip]{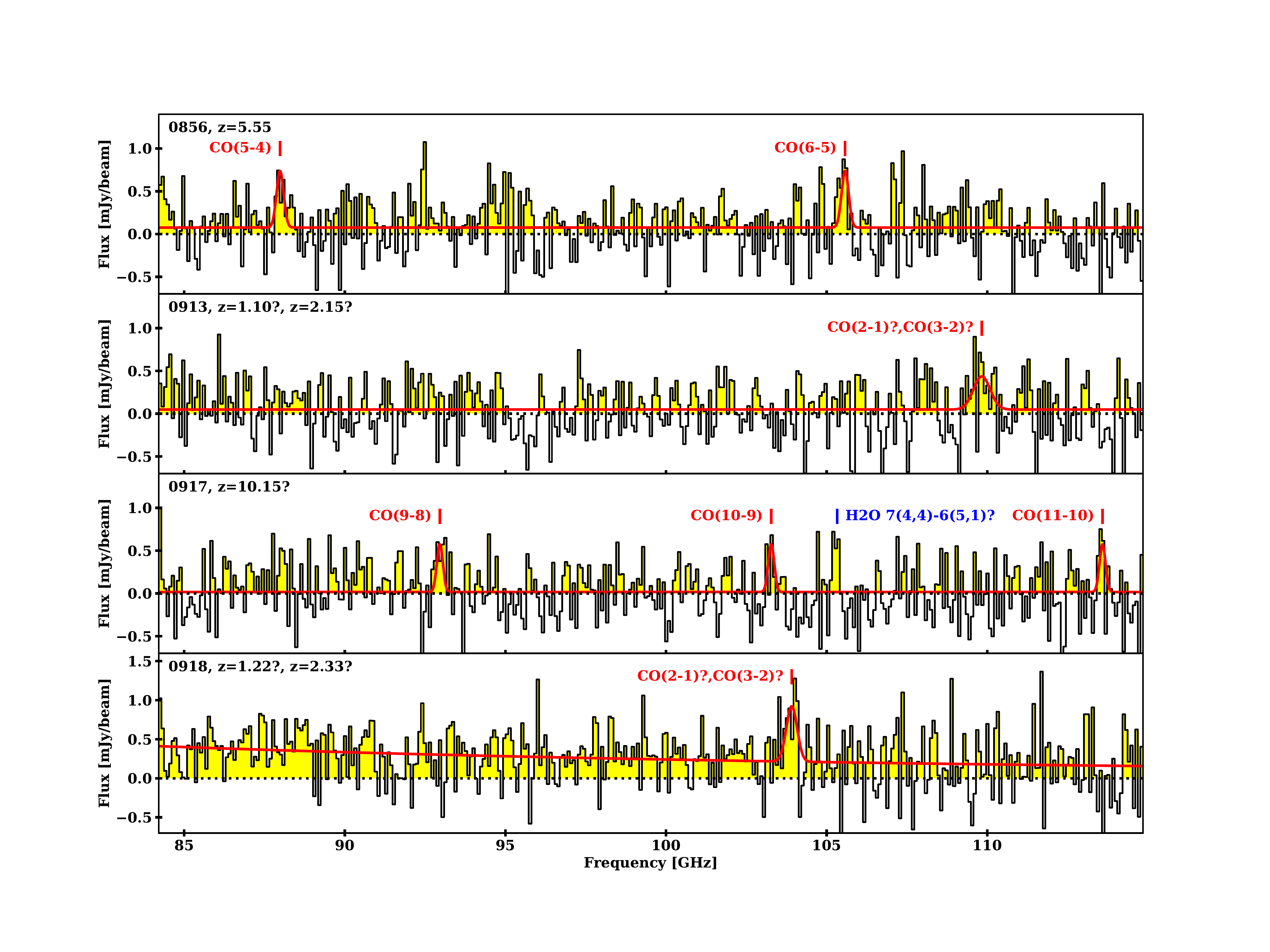} 
    \caption{The ALMA spectra with a 80\,MHz resolution for our four sources extracted from the positions of the hosts as seen in the $K_s-$band images. Table~\ref{tab:alma_spec} reports the line flux densities. We also present the fitting of a simple model for the observed lines. In the case of a possible redshift solution, we indicate the corresponding line transitions. See \S~\ref{sec:z_deter} for more details on the fitting and the redshift determination.}  
    \label{fig:ALMA_spec}
\end{figure*}

The $K_s$-band imaging was key to finding the host galaxies and we therefore extracted spectra over $0.8''$ radius centred on the $K_s$ positions. From this we subtracted a background spectrum created from a sky annulus (radius of $3.5-5.5''$). We present the spectra in Fig.~\ref{fig:ALMA_spec}. We then ran a 1D line-finder ({\tt sslf}\footnote{\url{https://github.com/cjordan/sslf}}) with a low threshold for line detection (f$_{\rm thres}>3\sigma$ and linewidth {\bf larger than} four channels). The only exception is 0917, where we used a lower threshold (f$_{\rm thres}>2.5\sigma$, see \S~\ref{sec:z_deter}). We found weak emission lines in these spectra {\bf and we verified they were also successfully detected in higher resolution spectra with 20, 35, 40, 60 and 75\,MHz-channel-width}. We describe the method for the redshift determination in \S~\ref{sec:z_deter}.

\begin{table*}
    \caption{ALMA spectra line measurements: detected and fitted. The detections have a central frequency, peak flux and width all with uncertainties. 
    The fitted measurements are too all detected lines simultaneously and 
    give a background continuum, amplitude, width and redshift (all with 
    uncertainties). 
    $^a$The line width here corresponds to $930\pm120$ km.s$^{-1}$ {\bf and $750\pm$200 km.s$^{-1}$} when converted into the rest-frame for 0856 and 0917, respectively. $^b$This line is only detected at very marginal level, note the width being lower than the spectral resolution (80MHz). $^c$This line is not fitted simultaneously with the CO lines.}
    \begin{center}
    \begin{small}
    \begin{tabular}{c cccc c ccccc}
    \hline
           & \multicolumn{4}{c}{Individual line detections with {\tt sslf}} & & \multicolumn{5}{c}{Results of simultaneous toy model line fit  (see \S~\ref{sec:z_deter})} \\
    \cline{2-5}
    \cline{7-11}
    Source & Frequency & Peak Flux & width & SNR & & line & cont. & amp. & width$^a$ & redshift \\
        & [GHz]            & [mJy] & [MHz] &    & &    & [$\mu$Jy] & [mJy] & [MHz] & \\
    \hline \hline
    \multirow{2}{*}{0856} & 88.21$\pm$0.05 & 0.64$\pm$0.23 & 129$\pm$53 & 3.2 & & CO(5-4) & \multirow{2}{*}{75$\pm$16} & \multirow{2}{*}{0.66$\pm$0.17} & \multirow{2}{*}{111$\pm$34} & \multirow{2}{*}{5.550$\pm$0.002} \\
    & 105.57$\pm$0.04 &  0.79$\pm$0.23 &  128$\pm$43 &  4.1 & & CO(6-5) & & & & \\
    \hline
    \multirow{2}{*}{0913} & 108.21$\pm$0.14 & 0.27$\pm$0.18 & 195$\pm$143 & 3.8 & & \multirow{2}{*}{-} & \multirow{2}{*}{-} & \multirow{2}{*}{-} & \multirow{2}{*}{-} & \multirow{2}{*}{-} \\
    & 109.89$\pm$0.13 & 0.35$\pm$0.16 & 245$\pm$126 & 4.0 & & & & & & \\
    \hline
    \multirow{4}{*}{0917} & 93.09$\pm$0.06 & 0.57$\pm$0.19 & 164$\pm$64 & 3.7 & & CO(9-8) & \multirow{3}{*}{16$\pm$16} & \multirow{3}{*}{0.56$\pm$0.15} & \multirow{3}{*}{94$\pm$30} & \multirow{4}{*}{10.154$\pm$0.003} \\
    & 103.41$\pm$0.08 & 0.37$\pm$0.23 & 108$\pm$80 & 2.8 & & CO(10-9)? & & & &  \\
    & 113.65$\pm$0.03 & 0.8$\pm$0.3 & 58$\pm$23 & 2.8 & & CO(11-10)$?^b$ & & & &  \\
    & 105.33$\pm$0.03 & 0.75$\pm$0.27 & 84$\pm$35 & 3.4 & & H$_2$O 7$_{4,4}$-6$_{5,1}$$?^{c}$ & - & - & - &  \\
    \hline
    0918 & 103.89$\pm$0.04 & 0.80$\pm$0.25 & 121$\pm$44 & 4.4 & & - & - & - & - & - \\
    \hline
    \end{tabular}
    \end{small}
    \end{center}
    \label{tab:alma_spec}
\end{table*}

\section{Results}

Due to the complex nature of our sources and our multi-wavelength dataset, we will present our results in the following order. First, we will review individual morphology from our continuum emission
(Figure~\ref{fig:morph}). Second, we will  present the spectral information through their SEDs. Third, we will determine the redshifts of our sources from their ALMA spectra and previously introduced information, and finally, we explore their NIR and radio properties. 

\subsection{Multi-wavelength morphologies}
\label{sec:morph}

\paragraph{GLEAM 0856:}
The $K_s-$band image reveals two sources. 
The resolved ALMA continuum (red contours) consists of an extended component (with a $>5\sigma$ peak) 
over $2-3''$.
Two very faint extensions eastward and southward are also detected at the $3\sigma$-level. 
 The ALMA emission is associated with
the eastern $K_s-$band source. We therefore consider this source to be the 
host of the radio galaxy, and extract our ALMA spectrum from this position. 
This interpretation is supported by weak radio emission ($2\sigma$) $2''$ north of the host galaxy and opposite the extended radio 
component. The stacked CO emission covers the host galaxy, but extends in the 
direction of the weak northern radio emission. 

\paragraph{GLEAM 0913:}

This radio source is the most extended of our sample in the radio (see Fig~\ref{fig:morph}). The ALMA continuum image presents three components with a total separation of $\sim35''$. These components appear to be related as the 19\,GHz ATCA morphology (magenta contours)  is consistent with a lobe-core-lobe morphology.

The identification of the host galaxy is straight forward as the central radio component is co-located with a bright source in the $K_s$ band (see the inset of Fig~\ref{fig:morph}). The corresponding NIR source is actually detected in the VIKING images, but was offset  $\sim 6''$ to the north of
our low-frequency centroid radio position (see purple circle in Fig.~\ref{fig:morph}). This demonstrates that higher resolution continuum data is essential to identify the host galaxy. We also show the moment 0 map of the ALMA spectral line at 109.89\,GHz as blue contours in the insets. 

\paragraph{GLEAM 0917:}

This source is the faintest of our sample in both NIR and radio. We see a faint continuum at 100\,GHz, detected at the 5$\sigma$-level. We also see two faint sources in the $K_s$ image with 
 the eastern one associated with the ALMA continuum detection which we identify as the host galaxy. 
The ALMA spectrum shows four weak lines, which we discuss in more details in S~\ref{sec:z_deter}. We present the average-stacked lines as the blue contours and see a detection centred on our object. 

\paragraph{GLEAM 0918:}

The $K_s$ image shows a bright source co-located with a double source in the ALMA continuum, likely corresponding to two synchrotron lobes. This host galaxy also has a detection in the VIKING data, but 
was misidentified due to the low-resolution radio data (note the large 
offset of $6''$ of the GLEAM position from this $K_s$ source). 
The higher resolution ALMA data has allowed us to unambiguously identify this source with the VIKING counterpart. Only one line is detected in the ALMA spectra, indicating likely a lower redshift source (as suggested by the strong VIKING detection). 

\begin{figure*}[t]
    \centering
    \begin{tabular}{cc}
        \includegraphics[width=0.48\textwidth]{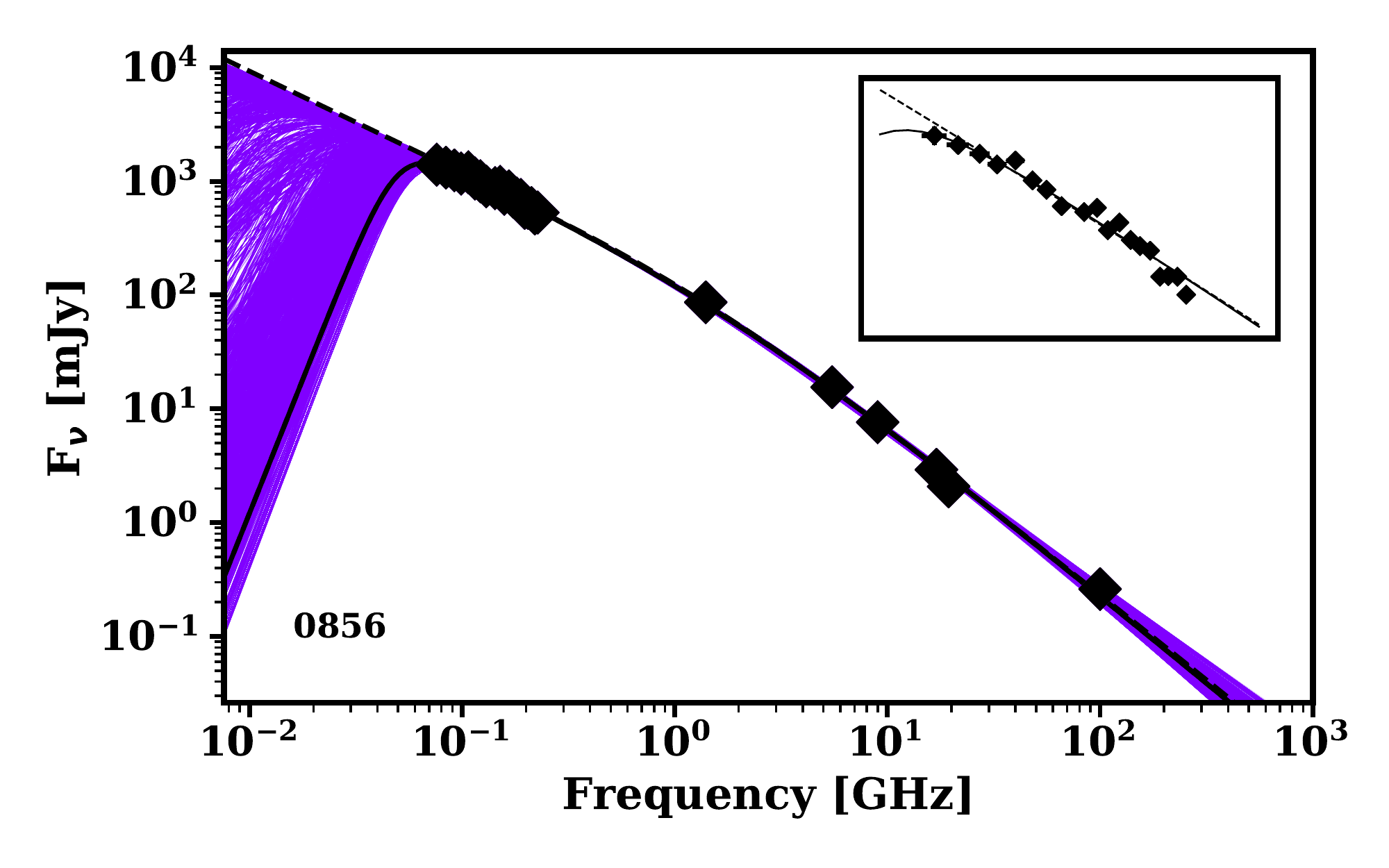} & \includegraphics[width=0.48\textwidth]{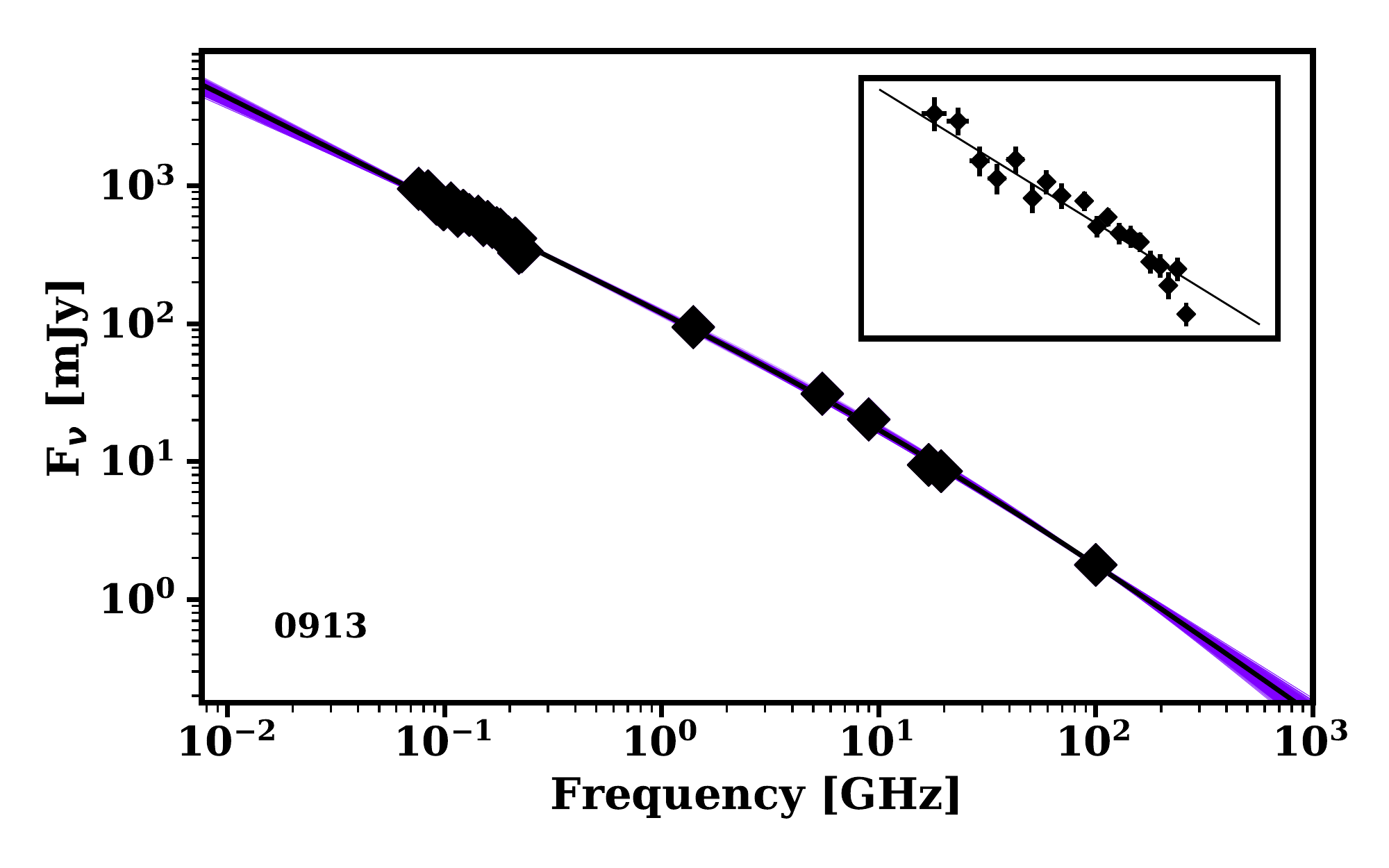} \\
        \includegraphics[width=0.48\textwidth]{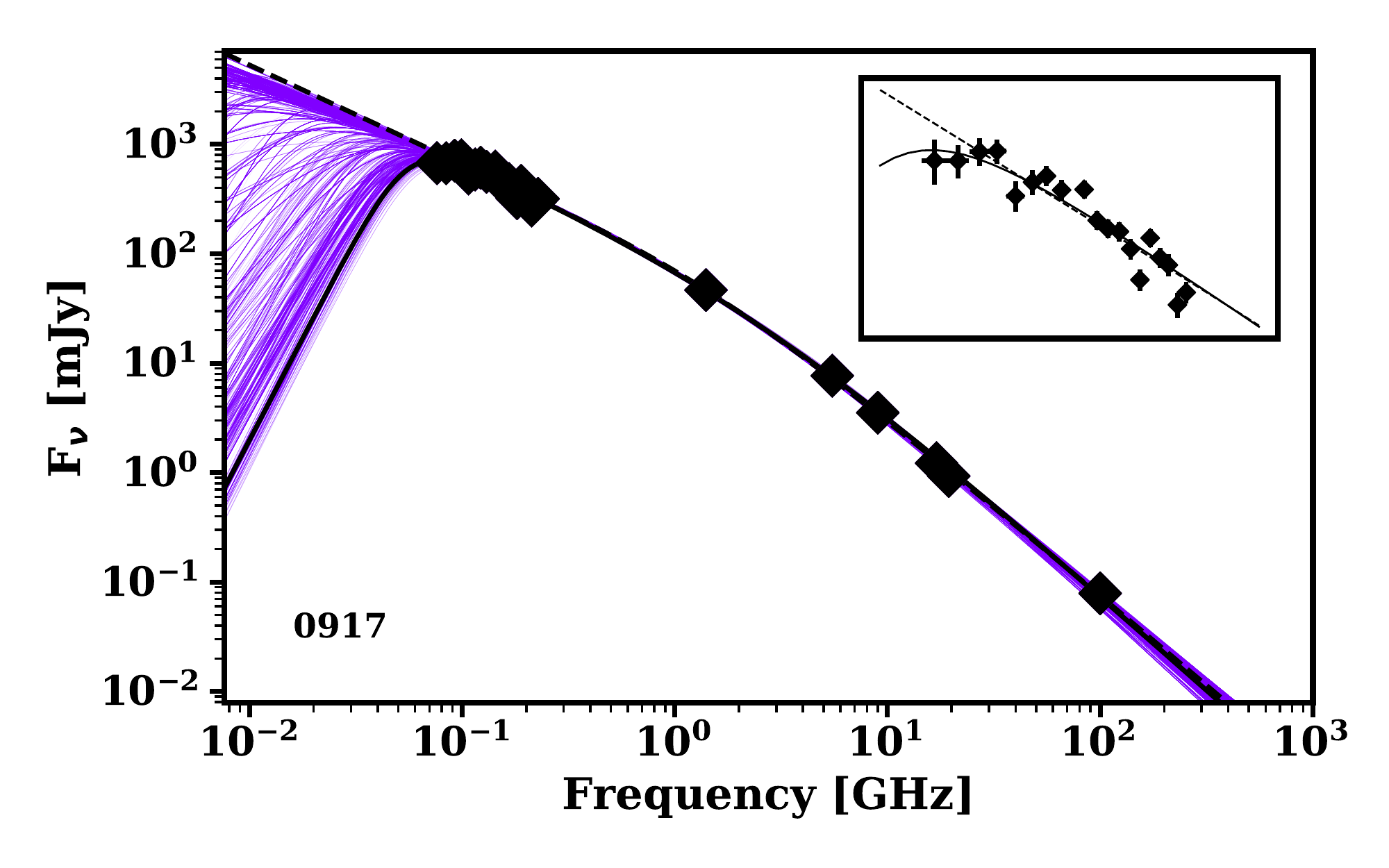} & \includegraphics[width=0.48\textwidth]{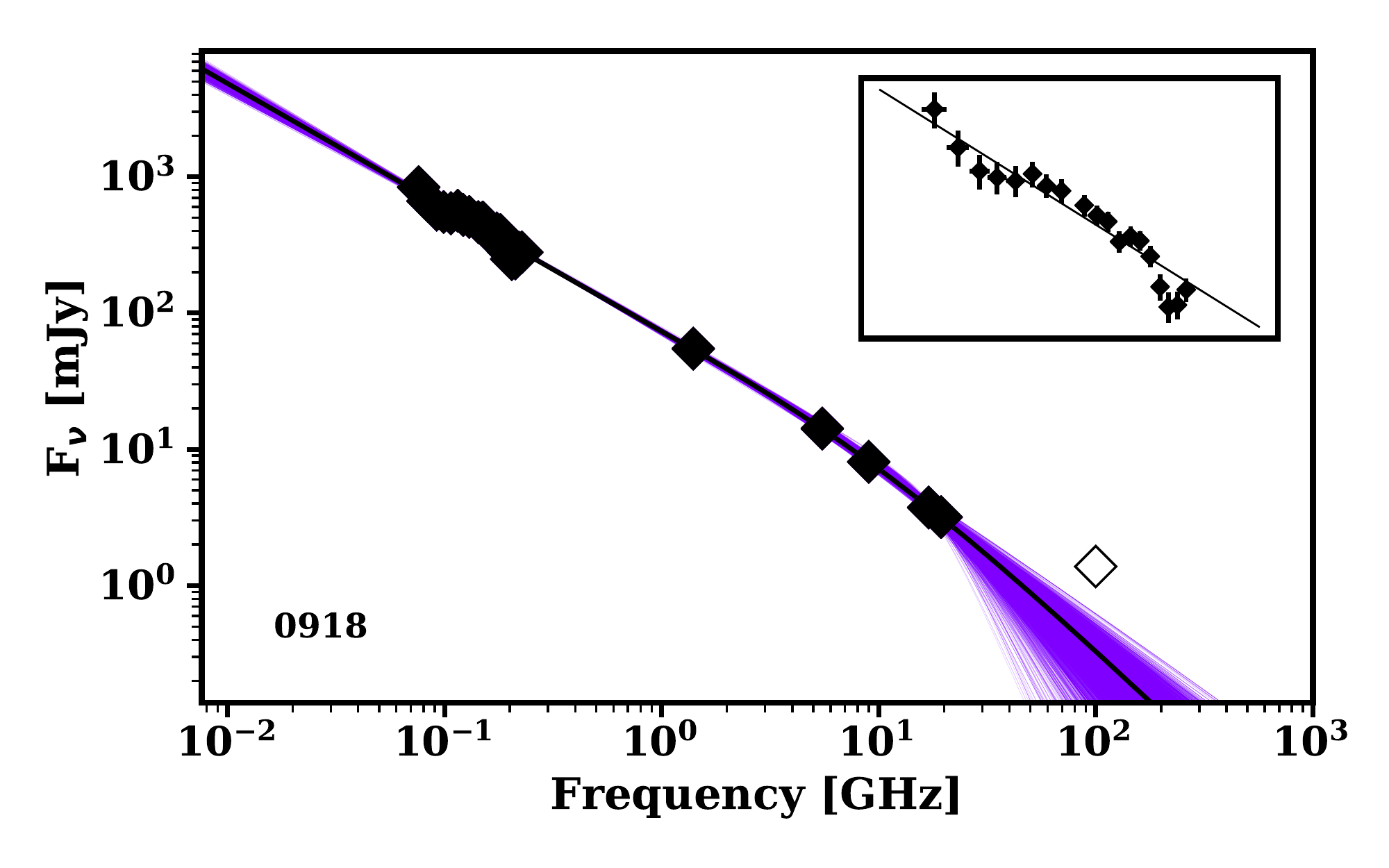} \\
    \end{tabular}
    \caption{The observed-frame radio SED for each source fitted with the triple power-law model for 0856 and 0917 and the double power-law model for 0913 and 0918 (all plotted as solid black lines). Uncertainties are
    plotted, but are hidden by the symbols. 
    The uncertainty is represented by the  scatter in the purple lines (see \S~\ref{sec:mrmoose}) 
    For 0856 and 0917, we also present the best fit for the double power-law as a black dashed-line. The data in each SED, with number of data points in parentheses, comprise, from low to high frequency, MWA (20), TGSS (1), NVSS (1), ATCA (4), ALMA (1). Note the open diamond for the ALMA data for 0918 is not included in our fit (see \S~\ref{sec:mrmoose}). Note that uncertainties are plotted, but are smaller than the symbol size. The insets are a zoom on the MWA data  with the best fit(s) from the wider SED overlaid.}
    \label{fig:sed_radio}
\end{figure*}

\subsection{Radio and optical SEDs}
\subsubsection{Radio broad-band SEDs}
\label{sec:mrmoose}

The complete radio SED, from 76\,MHz to 100\,GHz for each source is presented in Fig.~\ref{fig:sed_radio} and Table~\ref{tab:radio}. Note that an absolute calibration uncertainty is added in quadrature to each data-point uncertainty. We assume 8\% for GLEAM \citep{hurley-walker_galactic_2017}, 3\% for NVSS \citep{condon_nrao_1998} and 10\% for ALMA.

We perform SED fitting in order to characterise the break frequencies and change of spectral indices across the four decades in frequency for the total integrated emission. We use the SED fitting code {\tt MrMoose} \citep{drouart_mrmoose:_2018}, using three different models to fit the data: a simple power-law, a double power-law with a break frequency and a triple power-law with a double break frequency. We define the functions as follows\footnote{The more general formula is also available at \url{https://math.stackexchange.com/questions/2427089/how-do-i-smoothly-merge-two-power-laws}}:
\begin{eqnarray}
    S_\nu^{\rm spl}&=&N_{\rm spl}(\nu_{\rm obs})^{\alpha}, \\
    S_\nu^{\rm dpl}&=&N_{\rm dpl}(\nu_{\rm obs})^{\alpha_{\rm low}}\left(\frac{\nu_{\rm obs}}{\nu_{\rm b}}\right)^{\alpha_{\rm low}-\alpha_{\rm high}}, \\
    S_\nu^{\rm tpl}&=&N_{\rm tpl}(\nu_{\rm obs})^{\alpha_{\rm low}} \nonumber \\ 
    & & \times\left[1+\left(\frac{\nu_{\rm obs}}{\nu_{\rm b\_ low}}\right)^{|\alpha_{\rm low}-\alpha_{\rm med}|}\right]^{\rm sgn(\alpha_{\rm med}-\alpha_{low})}  \nonumber \\
    & & \times\left[1+\left(\frac{\nu_{\rm obs}}{\nu_{\rm b\_high}}\right)^{|\alpha_{\rm med}-\alpha_{\rm high}|}\right]^{\rm sgn(\alpha_{\rm high}-\alpha_{\rm med})}
\end{eqnarray}

where $\nu_{\rm obs}$ is the observed frequency, $\nu_{\rm b}$, $\nu_{\rm b\_low}$ and $\nu_{\rm b\_high}$  the break frequencies, $\alpha$, $\alpha_{\rm low}$, $\alpha_{\rm med}$ and $\alpha_{\rm high}$, the spectral indexes, and $sgn$ referring to the sign of the operation. Note the absolute value for the spectral index difference.

\begin{table*}[t]
    \caption{Results from the observed-frame radio SED fitting. We report the best fit parameter for each model (single, double and triple power-law from top to bottom) along with their uncertainties defined as the 25 and 75 percentiles. The break frequencies are given as the log of the frequency in Hz, ln$(L)$ refers to the log of the maximum of the likelihood function used to calculate the AICc criteria (see Eq.~\ref{eq:aicc}).}
    \centering
    \begin{tabular}{ll cccc}
        \hline
        Model & Parameter &  0856 & 0913 & 0917 & 0918 \\
        \hline \hline
        & $N_{\rm spl}$ & $-13.48_{-0.02}^{+0.02}$ & $-16.12_{-0.01}^{+0.02}$ & $-13.10_{-0.03}^{+0.03}$ & $-15.51_{-0.03}^{+0.04}$ \\
        single power-law & $\alpha$ & $-1.18_{-0.00}^{+0.00}$ & $-0.87_{-0.00}^{+0.00}$ & $-1.25_{-0.00}^{+0.00}$ & $-0.96_{-0.00}^{+0.00}$ \\
        & ln$(L)$ & -282.04 & -74.89 & -297.31 & -48.52 \\
        & AICc & 568.08 & 153.78 & 598.62 & 101.04 \\
        \hline
        &$N_{\rm dpl}$ & $-24.16_{-0.10}^{+0.11}$ & $-25.16_{-0.15}^{+0.35}$ & $-24.42_{-0.09}^{+0.09}$ & $-25.30_{-0.07}^{+0.09}$ \\
        &$\nu_{\rm b}$ & $9.38_{-0.10}^{+0.08}$ & $10.62_{-0.36}^{+0.15}$ & $9.37_{-0.08}^{+0.07}$ & $10.28_{-0.07}^{+0.06}$ \\
        double power-law &$\alpha_{\rm low}$ & $-0.89_{-0.02}^{+0.02}$ & $-0.78_{-0.02}^{+0.02}$ & $-0.91_{-0.02}^{+0.03}$ & $-0.91_{-0.01}^{+0.01}$ \\
        &$\alpha_{\rm high}$ & $-1.51_{-0.02}^{+0.03}$ & $-1.14_{-0.05}^{+0.19}$ & $-1.65_{-0.02}^{+0.03}$ & $-1.59_{-0.17}^{+0.17}$ \\
        &ln$(L)$ & -49.62 & -33.51 & -45.54 & -28.49 \\
        &AICc & 108.23 & 76.01 & 100.08 & 65.98 \\
        \hline
        &$N_{\rm tpl}$ & $-17.06_{-2.27}^{+2.35}$ & $-14.48_{-3.74}^{+5.05}$ & $-18.07_{-2.45}^{+1.55}$ & $-38.74_{-8.32}^{+10.71}$ \\
        &$\nu_{\rm b\_low}$ & $6.91_{-0.92}^{+0.74}$ & $6.14_{-0.57}^{+0.63}$ & $7.22_{-0.65}^{+0.38}$ & $6.74_{-0.38}^{+0.30}$ \\
        &$\nu_{\rm b\_high}$  & $9.11_{-0.13}^{+0.21}$ & $10.85_{-0.32}^{+0.22}$ & $9.21_{-0.12}^{+0.12}$ & $10.42_{-0.05}^{+0.06}$ \\
        triple power-law &$\alpha_{\rm low}$ & $2.17_{-1.46}^{+1.57}$ & $1.82_{-1.11}^{+1.04}$ & $2.35_{-1.63}^{+0.95}$ & $2.51_{-1.57}^{+1.18}$ \\
        &$\alpha_{\rm med}$ & $-0.78_{-0.07}^{+0.04}$ & $-0.76_{-0.02}^{+0.03}$ & $-0.82_{-0.05}^{+0.05}$ & $-0.91_{-0.01}^{+0.02}$ \\
        &$\alpha_{\rm high}$ & $-1.51_{-0.04}^{+0.04}$ & $-1.32_{-0.11}^{+0.09}$ & $-1.69_{-0.04}^{+0.04}$ & $-2.18_{-0.50}^{+0.35}$ \\
        &ln$(L)$ & -46.78 & -34.32 & -42.37 & -28.88 \\
        &AICc & 109.55 & 84.65 & 100.75 & 73.75 \\
         \hline
    \end{tabular}
    \label{tab:sed_results}
\end{table*}

We are interested here in 
the SED from the total flux emission (sum of all components, see Table~\ref{tab:radio}), and therefore only focus on  models with differing numbers of power-laws. Characterising the physical processes at work further is beyond the scope of this paper, especially given that only one source has a definite redshift (see \S~\ref{sec:z_deter})
We compare the relative likelihood for each model with the Akaike Information Criteria \citep[AICc][]{akaike_new_1974,burnham_model_2002} which is defined as:

\begin{equation}
\label{eq:aicc}
{\rm AICc} = 2k-2{\rm ln}(L) + \frac{2k(k+1)}{(n-k-1)},
\end{equation}

with $n$ the number of data points, $k$ the number of free parameters and ln$(L)$ the maximum of the likelihood function (calculated with {\tt MrMoose}). Note the first term, $2k$, penalising the addition of free parameter and the last term an added correction in the case of small sample sizes (this can be seen as an extra penalty for an increased number of free parameters on small samples). We report the results of the fitting in Table~\ref{tab:sed_results} and plot the prefered model in Fig.~\ref{fig:sed_radio}.

The AICc criteria indicates that the model which best represents the
sources overall is the double power-law model. Moreover, it is clear that the triple power-law, while providing a good fit, does not improve significantly the AICc for 0918 and 0913 even with the curvature in the MWA frequency bands. As for 0856 and 0917, the AICc scores are similar for the double and triple power-law models which indicate that both models are equally preferred. We note that the best fit for the triple power-law reproduces the curvature in the MWA data, albeit with a large uncertainty on the spectral slope at lower frequency, $\nu<70\,$MHz,  where the fit is unconstrained. For all sources, we measure  a double power-law break frequency in the $2-42\,$GHz range (note how the ATCA data provides for a strong constraint here). The lower frequency spectral index is moderately steep with $\alpha<$-0.7 and in reasonable agreement with the GLEAM-only spectral index from our polynomial fit used in \S~\ref{sec:cand_selec}. The higher frequency spectral index is systematically steeper by $\Delta(\alpha)=0.36-0.7$.

In the case of 0918, the ALMA continuum point is not included  in the model fitting as it would imply an up-turn/flattening of the radio SED or a separate component. This could be due to either a separate synchrotron component with a very high frequency turn-over or possible inverse Compton emission from the lobes. Without further data it is impossible to model either possibility.

\subsubsection{Broad band optical-NIR SEDs}
\label{sec:photz}

We note that two sources (GLEAM 0913 and 0918) are detected in  all VIKING NIR bands due to mis-identification arising from the low-frequency, low-resolution radio data (see \S~\ref{sec:morph}). Having access to a significant part of the SED, we use the photometric redshift code EAZY with default settings \citep{brammer_eazy:_2008} to estimate the redshift of these two sources. While presenting large uncertainties, these redshift are useful when compared with the ALMA spectral line redshift solutions (see next sub-section). We obtain $z_{\rm 0913}$=0.96$^{+0.17}_{-0.12}$ and $z_{\rm 0913}$=0.77$^{+0.22}_{-0.26}$ with 68\% confidence limits. Of the two other sources, 0856 is weakly detected in $z-$band and 0917 is not detected in any VIKING band. We can obtain some information on the galaxy host using the $K-z$ relation (see Fig.~\ref{fig:Kz} and \S~\ref{sec:kz}). 

\subsection{Redshift determination from the ALMA spectra} 
\label{sec:z_deter}

Given the spectral scan in Band 3 (84-115\,GHz), we have three possibilities:
\begin{itemize}
\item no line is detected in this range and we are either observing a source located at a redshift with no CO line in this range (at $0.36<z<1.0$ and $1.74<z<2.0$) or a source with molecular lines too faint to be detected,
\item a single line is detected and the redshift solution is ambiguous but constraints may be derived from the absence of other molecular lines or,
\item two lines or more are detected across the spectra and the redshift can be unambiguously measured and/or any interloper can be clearly identified. 
\end{itemize}

This is the method used by the SPT team to determine the redshifts of strongly lensed sub-mm galaxies \citep{weis_alma_2013,strandet_redshift_2016}. Note that while working extraordinary well on lensed sources thanks to their compactness and magnification, using this technique on un-lensed sources is far more challenging. We designed our sample selection to find multiple lines (if at $z>3.0$), i.e. falling into option (iii) above and therefore obtaining an unambiguous redshift measure.

\label{sec:alma_spec_lines}
\begin{figure*}
    \centering
    \includegraphics[width=1.0\textwidth,trim=0 20 0 20,clip]{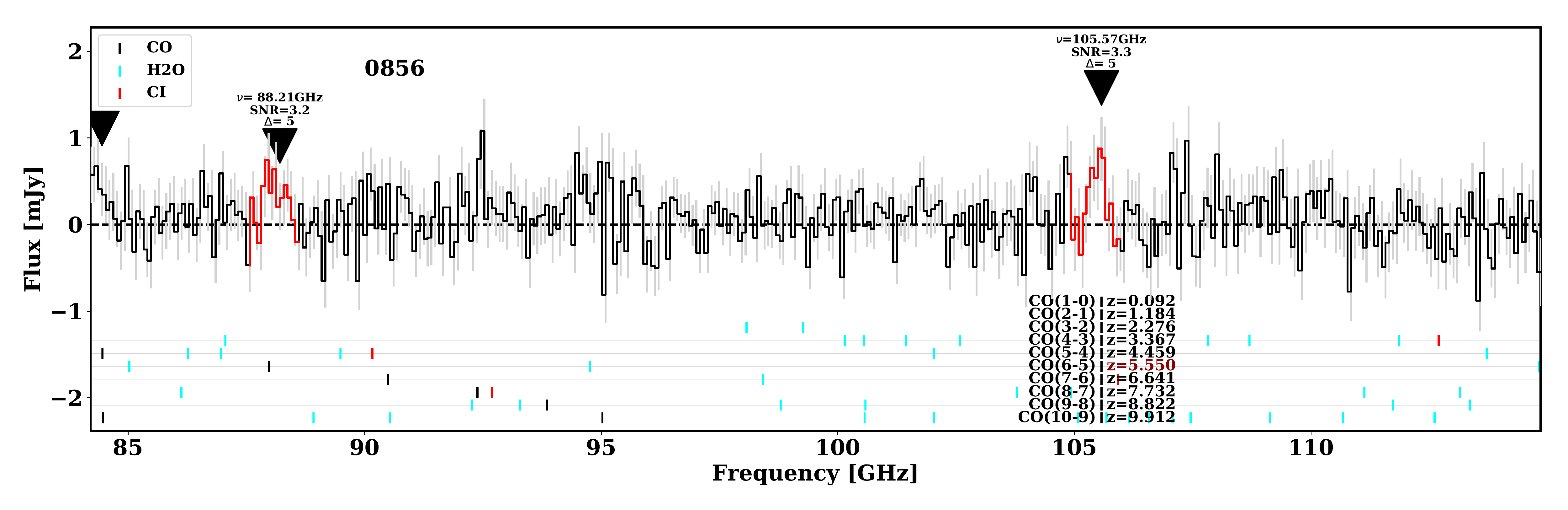}
    \includegraphics[width=1.0\textwidth,trim=0 20 0 20,clip]{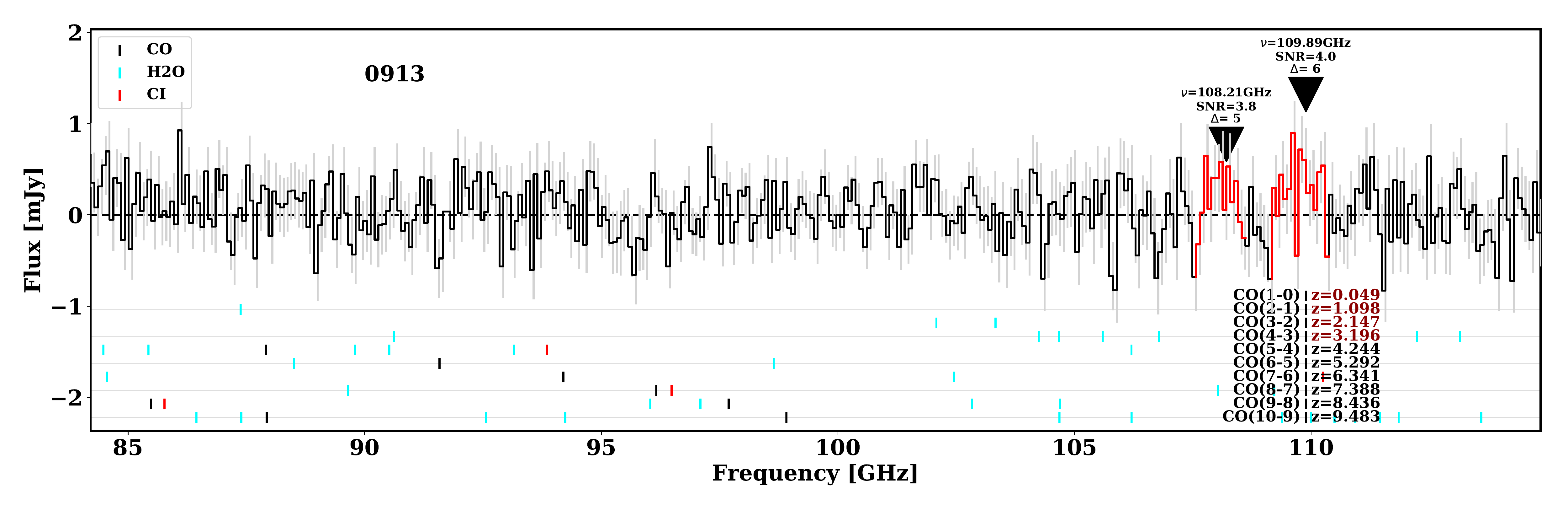}
    \includegraphics[width=1.0\textwidth,trim=0 20 0 20,clip]{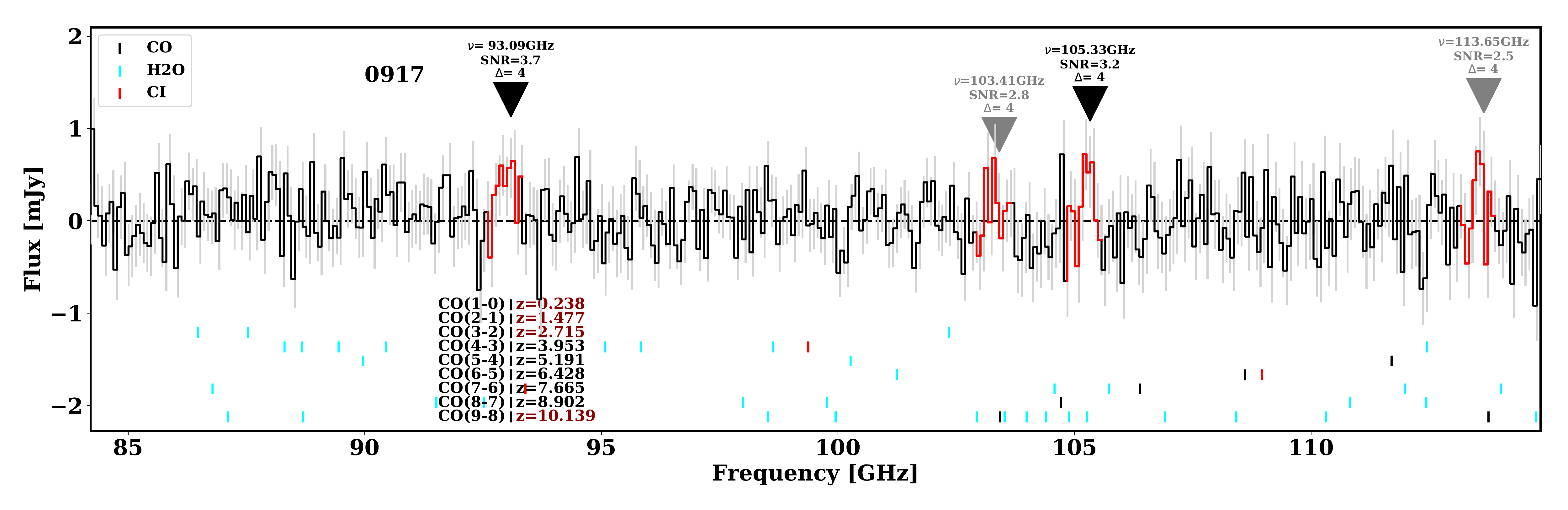}
    \includegraphics[width=1.0\textwidth,trim=0 20 0 20,clip]{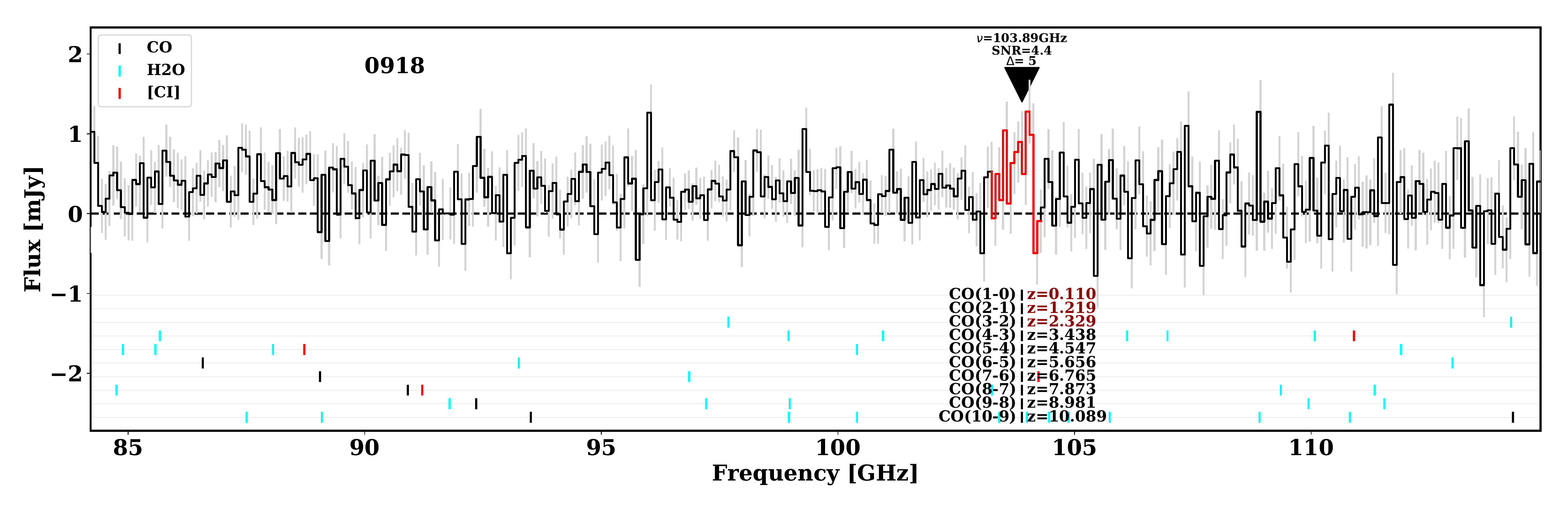}
    \caption{Redshift determination for our four sources from the ALMA spectra. This figure shows the 1D spectra extracted at the position indicated in Fig.~\ref{fig:morph} and presented in Fig.~\ref{fig:ALMA_spec}. The grey error-bars are showing the rms per channel. The downward triangles and the red part of the spectra refer to the detection of the lines by {\tt sslf} (see \S~\ref{sec:data_alma}). The line characteristics are reported above the markers (central frequency, signal-to-noise ratio and FWHM in number of channels). We highlighted the detections in the 2.5-3$\sigma$ range in grey for 0917. Taking the highest signal-to-noise line and assuming this is a CO line, the vertical black markers are reporting the combination of the possible redshift along with the prediction of the other transition CO lines, as well as [CI] and H$_2$O. We highlight the potential redshift solutions in dark red (see \S~\ref{sec:z_deter}). }
    \label{fig:z_deter}
\end{figure*}

We examine our lines sequentially in  decreasing signal-to-noise order, assuming that we are observing a CO line and therefore predict the frequency of other molecular lines, CO, [CI] and H$_2$O (using splatalogue\footnote{\url{https://www.cv.nrao.edu/php/splat/}}) for the corresponding redshifts (the vertical bars in Fig.~\ref{fig:z_deter}). We do not consider the HCN and HCO+ lines as they are expected to be much fainter. With even a single line identification, we can exclude  certain redshift solutions given that other lines should be detected in our broad frequency range. In the case of multiple line detection, we compare the estimated frequencies with the other detected lines with {\tt sslf}. Our approach is summarised in Fig.~\ref{fig:z_deter}, where a good match is usually obvious (e.g. 0856). Thus, we perform a fitting on the spectra using this redshift prior, adding the corresponding number of Gaussians required to reproduce the spectra (note that we assume the exact same amplitude and width for all lines). We also add a continuum component as a constant across the frequency range (excepted 0918 where a spectral index is added as a strong gradient is observed in the spectra, see Fig.~\ref{fig:ALMA_spec} and Table~\ref{tab:alma_spec}). We now review our  sources individually, by increasing  order of complexity.

\paragraph{GLEAM 0918: }

A single detected line is centered at 103.89\,GHz which would translate to $z_{\rm CO(1-0)}$=0.110, $z_{\rm CO(2-1)}$=1.219, $z_{\rm CO(3-2)}$=2.329, $z_{\rm CO(4-3)}$=3.438 and $z_{\rm CO(5-4)}$=4.547 (see Fig.~\ref{fig:z_deter}). We can discard the $z$=3.438 and $z$=4.547 solutions as the [CI](1-0) should appear in our frequency range. Conversely, we also can discard the lowest redshift solution ($z$=0.110) as the related radio luminosity (see Fig.~\ref{fig:radio_lum}), $K_s$ luminosity (see Fig.~\ref{fig:Kz}) and the predicted CO flux (see Fig.~\ref{fig:CO_prediction}) are not compatible for such a low redshift source. Hence, this source is at $z=1.219$ or $z=2.329$. Given the result of the photometric redshift (see \S~\ref{sec:photz}), the first solution appears 
the more likely. 

\paragraph{GLEAM 0913: }

The two lines detected in 0913 are too close to each other to come from two different CO transitions (and too far apart to have originated from the same source). Overlaying the contours of the two lines on the $K_s$-band image shows that the 108.21\,GHz line is very likely spurious. We therefore do not consider this line anymore. From the 109.89\,GHz line, the redshift solutions are $z_{\rm CO(1-0)}$=0.049, $z_{\rm CO(2-1)}$=1.098, $z_{\rm CO(3-2)}$=2.147 and $z_{\rm CO(4-3)}$=3.196. Any higher in redshift and we would see a second line in our frequency coverage (either [CI] or a CO transition, see Fig.~\ref{fig:z_deter}). For the same reason as for 0918, we can discard the lowest redshift solution. When comparing these redshift solutions with the phot-z estimate (Fig.~\ref{fig:sed_gama}), the solution at $z=1.098$ appears most probable. Additionally no radio source of this size has been observed above $z\sim 2.5$ \citep{de_breuck_spitzer_2010}. 

\paragraph{GLEAM 0856: }

This sources is the prototype source that our selection method is designed to find. Two lines are detected at 88.21\,GHz and 105.57\,GHz. Assuming they are both CO, the redshift is unambiguously determined  from the fit as $z=5.550\pm0.002$ (Table~\ref{tab:alma_spec}). %Moreover, this redshift is also consistent with the weak detection in the $z$ band in the VIKING detection, where the Ly$\alpha$ line would be redshifted at $0.79\,\mu$m (Fig.~\ref{fig:sed_gama}). 

\subsubsection{The complex case of GLEAM 0917}

GLEAM 0917 is the most complex spectrum to interpret. If considering only the two lines detected above 3$\sigma$ at 93.08\,GHz and 105.33\,GHz, these detections do not provide any redshift solution for a single source (see Fig.~\ref{fig:z_deter}). 
Alone, the 93.08\,GHz line would correspond to $z_{\rm CO(2-1)}$=1.477 and $z_{\rm CO(3-2)}$=2.745 (we discard the lower redshift CO(1-0) solution as for 0918). However, while not impossible, these redshift solutions present some problems with our NIR SED. The $K_s$ magnitude would point to a fainter system. Considering the 105.33\,GHz line on its own gives the solutions of $z_{\rm CO(2-1)}$=1.189 and $z_{\rm CO(3-2)}$=2.283. By decreasing the line detection threshold of {\tt sslf} to 2.5$\sigma$, two other lines are detected at 103.41\,GHz and 113.65\,GHz (indicated by grey triangles in Fig.~\ref{fig:z_deter}). The addition of these two lines along with the 93.08\,GHz line does provide a unique redshift solution at $z$=10.154$\pm$0.003 for CO(9-8), CO(10-9) and CO(11-10). However, this possibility comes with various caveats.

Firstly, the width of the lines are not similar; they become narrower with increasing frequency. In particular, the highest frequency line is best fit with a width on par with the one tapered $80\,$MHz bin. We therefore applied the same procedure to our smaller channel-width cubes. The three lines are still detected and still become narrower with increasing frequency. Part of this effect could originate from the noise increase due to the increasing atmospheric opacity (relatively strong toward the end of the band at $\nu>110\,$GHz in bad weather conditions). 

Secondly, to exclude the possibility that these lines are spurious we generated simulated spectra with the same Gaussian noise rms, but no true signal, and run {\tt sslf} with the exact same parameters ($f_{\rm thres}=2.5\sigma$ and linewidth {\bf larger than} four channels). Assuming a conservative frequency difference between the central frequency of the different CO transition lines to be a shift of $<3$ channel bins with respect to each other (corresponding to a significant shift up to 240\,MHz), out of $10^6$ simulated spectra only 626 cases (corresponding to a 0.06\% chance) are able to reproduce a spectra with three detected ``lines'' at $>2.5\sigma$. Note that this matching makes neither an  assumption on the line width (except being {\bf larger than} four channels to be detected by {\tt sslf}) nor on the amplitude of the lines. Any further constraints on these parameters will result in an even lower probability. We also note that all physical quantities (described in the following sections), the $K_s$ magnitude (Fig.~\ref{fig:Kz}), the 500\,MHz and 3\,GHz luminosities (Fig.~\ref{fig:radio_lum}) and the CO properties (Fig.~\ref{fig:CO_prediction}) are consistent with a powerful radio loud AGN at such an extreme redshift. 

 Thirdly, the $z=$10.15 solution would suggest that we are possibly observing a water line at 105.33\,GHz. The emission of H$_2$O is known to be rather complex to trace given the numerous excitation states available \citep{werf_water_2011}. This line is nonetheless routinely detected at high redshift \citep{weis_alma_2013,wang_alma_2013,gullberg_alma_2016}. \cite{wilson_stacked_2017} combined the sub-mm spectra of a large number of star-forming sources in the 0.1$<z<$4 range. The closest in frequency for this given redshift is the H$_2$O 7$_{4,4}$-6$_{5,1}$ transition ($\nu_{\rm rest}$=1172.526\,GHz) and is blue-shifted compared to the systemic redshift by $\sim300\,$km\,s$^{-1}$.  However, this line has not been detected at any redshift so far and the condition required to emit such a line would mean we would observe other lines in the full frequency range. The closest observed water line transition is the H$_2$O 3$_{2,1}$-3$_{1,2}$ \citep[at $\nu_{\rm rest}=1162.911\,$GHz detected by][]{wilson_stacked_2017}. However, this corresponds to a $\sim$2500\,km\,s$^{-1}$ blueshift, which appears to be very unlikely. We therefore conclude that this line might be from a different object along the line-of-sight,  or perhaps more likely, a spurious detection.

\subsection{Interpretation}

\subsubsection{NIR Observations and,the {\it K}-{\it z} Relation}
\label{sec:kz}

\begin{figure}
    \centering
    \includegraphics[width=0.5\textwidth,trim= 20 20 0 0,clip]{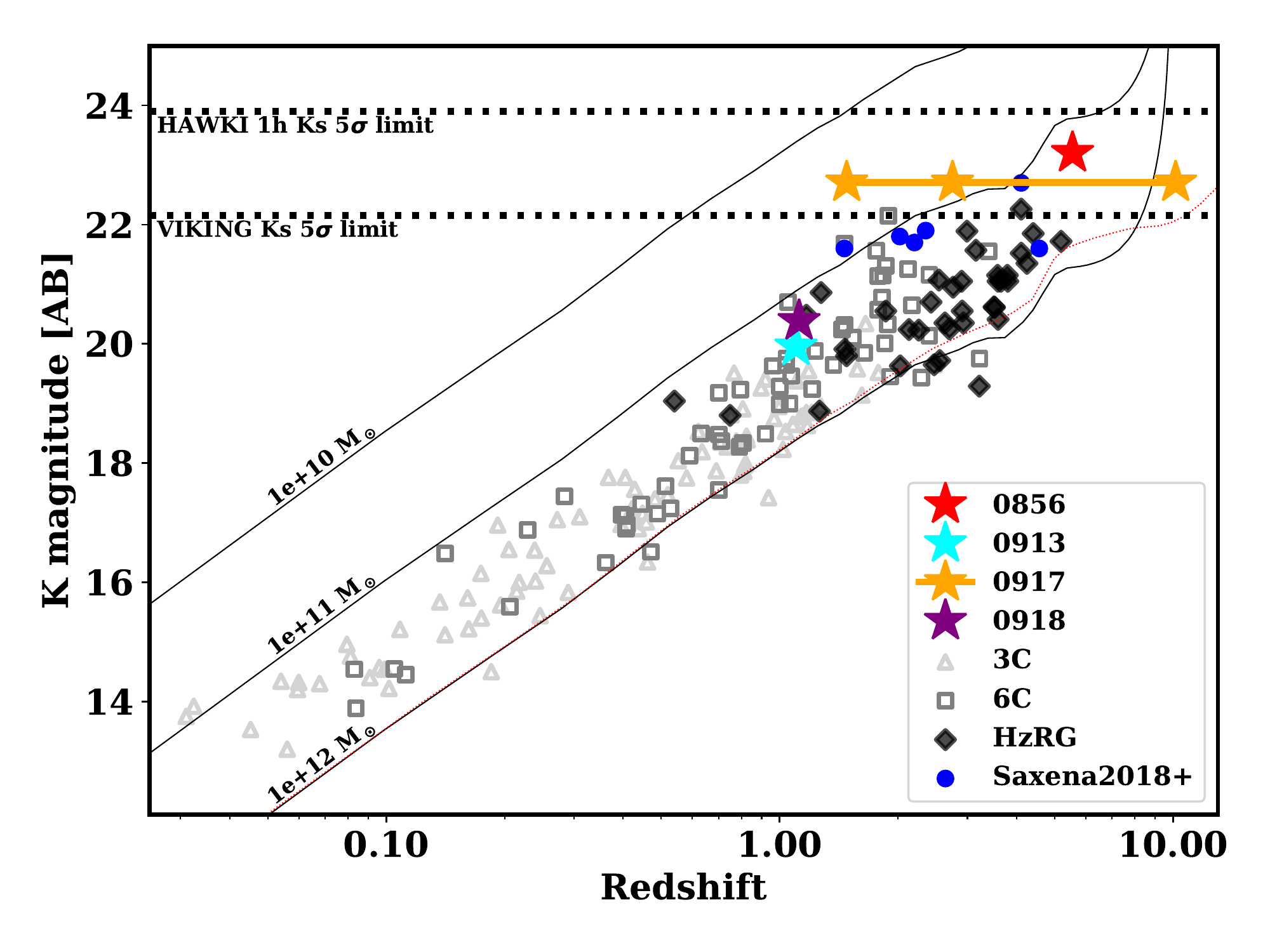}
    \caption{$K-z$ relation diagram showing the observed $K-$band magnitude against the redshift of known powerful radio galaxies (see legend). Our four candidates are represented with coloured stars and with lines connecting possible redshift solutions from a combined analysis of all the data in hand. The VIKING and HAWK-I detection limits are indicated by dotted lines. The tracks correspond to the elliptical templates from P\'EGASE.2 \citep{fioc_pegase:_1997} scaled to reported stellar masses.}
    \label{fig:Kz}
\end{figure}

The $K_s$ magnitudes provide us with our first insight of the properties of the sources, as well as helping to solve the case of degenerate redshift solutions as presented in \S~\ref{sec:z_deter}. Fig.~\ref{fig:Kz} shows our sources along with samples of powerful radio galaxies samples \citep[3C, 6C and other HzRGs;][]{lilly_stellar_1984,eales_first_1997,van_breugel_morphological_1998}. The $K-z$ relation shows that powerful radio galaxies form a correlation in observed $K-$band and redshift space which is modelled as being due to emission from massive galaxies \citep[$M_*\sim 10^{11-12}\,M_{\odot}$,][]{rocca-volmerange_radio_2004}.  The classic interpretation of this result is that the most luminous radio galaxies are powered by the most massive black holes lying in the most massive galaxies. Evidence for the large black hole masses comes from  \cite{nesvadba_evidence_2008} and \cite{drouart_rapidly_2014}. Hence, from this diagram, and as mentioned previously (\S~\ref{sec:alma_spec_lines}), the most likely redshift solutions for 0913 and 0918 appear to be consistent with the main relation. The $K-z$ relation arguably becomes less certain at the redshifts we aim to probe here ($z>5$) as the observed $K$-band shifts from the blue optical to the ultra-violet rest-frame. {\bf  At these wavelengths, the star-formation properties (star formation rate, age of the stellar population and dust content) can account for significant differences to the emission.} We see that the redshift solutions for 0917 are offset from the main relation. However, no powerful radio galaxy has been detected with $K\le 21.5$ at $z>5$.

\subsubsection{Radio luminosity interpretation}

\begin{figure}[t]
    \centering
    \includegraphics[width=0.5\textwidth,trim= 20 20 0 0,clip]{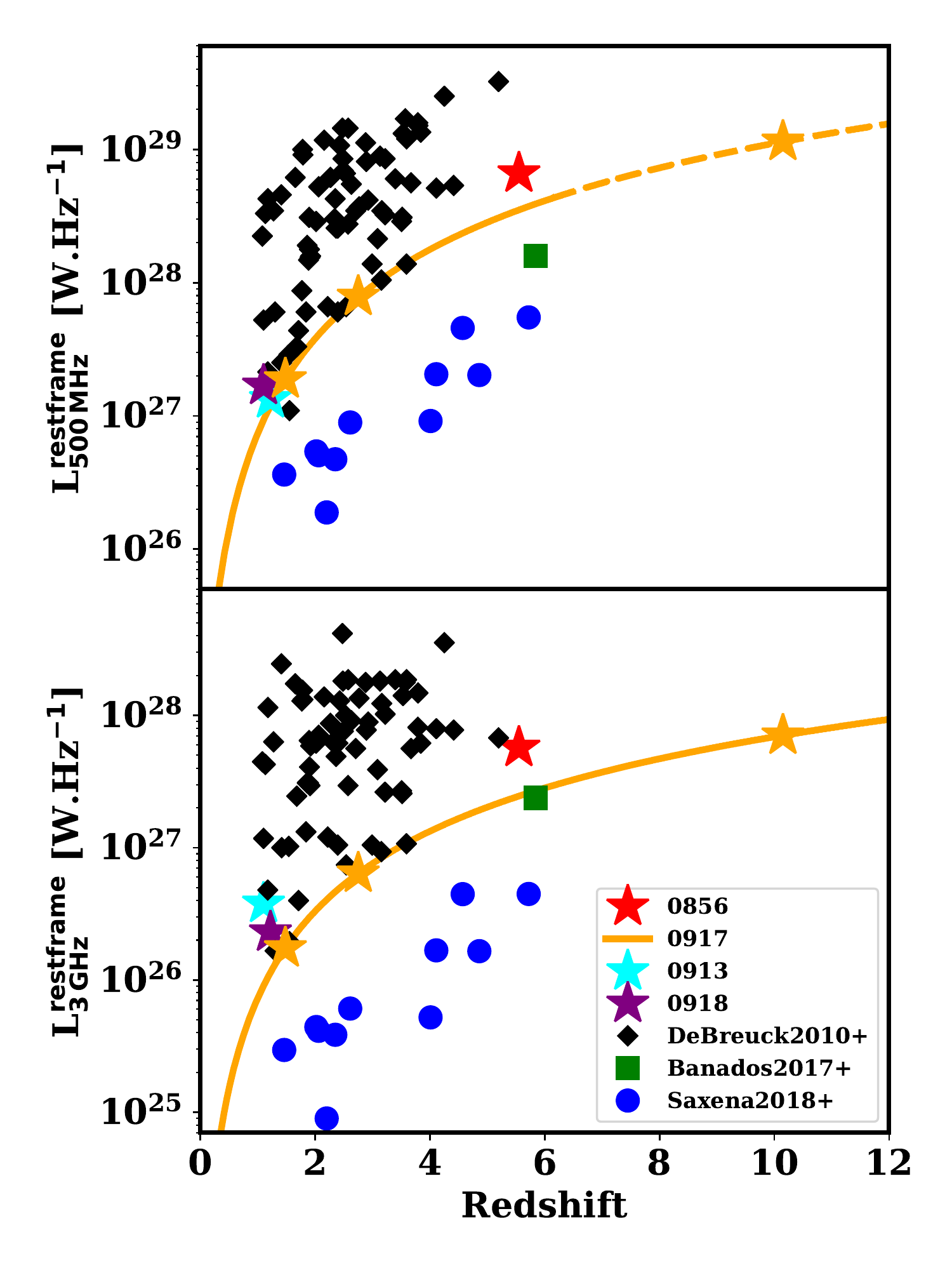}
    \caption{Radio luminosity at rest-frame 500\,MHz (top) and 3\,GHz (bottom)  plotted against redshift of samples of powerful high redshift radio galaxies. We present our four objects as stars with a line connecting multiple redshift solutions. 
    The solid lines are luminosities determined from the SED within our frequency coverage and the dashed line is where an extrapolation is required  from our best SED fitting (below the GLEAM limit, at $<$70\,MHz). We also plot the sample from \cite{saxena_nature_2019}, and the quasar from \cite{banados_powerful_2018}, recalculated at the relevant frequency from the flux densities and spectral indexes provided.}
    \label{fig:radio_lum}
\end{figure}

Having access to the radio SED from 70\,MHz to 115\,GHz, one can determine the rest-frame 500\,MHz and 3\,GHz luminosities. We show the distribution of these luminosities with redshift in Fig.~\ref{fig:radio_lum} from the SEDs previously shown in Fig.~\ref{fig:sed_radio}. Note that the lowest frequency GLEAM data point (70\,MHz) allows us to determine the 500\,MHz luminosity from data only up to $z=6.14$. After that we must extrapolate (via our best fit SED in our case) to estimate luminosities at higher redshift. The 3\,GHz rest-frame luminosity is not affected by this observational limit.

Fig.~\ref{fig:radio_lum} shows 0913 and 0918 have comparable radio luminosities to the bulk of previous samples of high-z powerful radio galaxies. In the case of 0856, the source also appear very similar to the HeRG\'E sample \citep{de_breuck_spitzer_2010}, but at higher redshift. For 0917, the luminosity at $z$=10.15 would be consistent with a very bright source, but nothing abnormally bright when compared with a similar sample at lower redshift. We also note that the break frequency for the four objects differ significantly  for the double power-law model. The higher redshift objects (GLEAM 0856 and potentially GLEAM 0917) have lower frequency breaks, consistent with them lying at higher redshift.

\subsubsection{Molecular gas properties and predictions}

\begin{figure*}[th]
    \centering
    \includegraphics[width=1\textwidth,trim=0 20 0 0,clip]{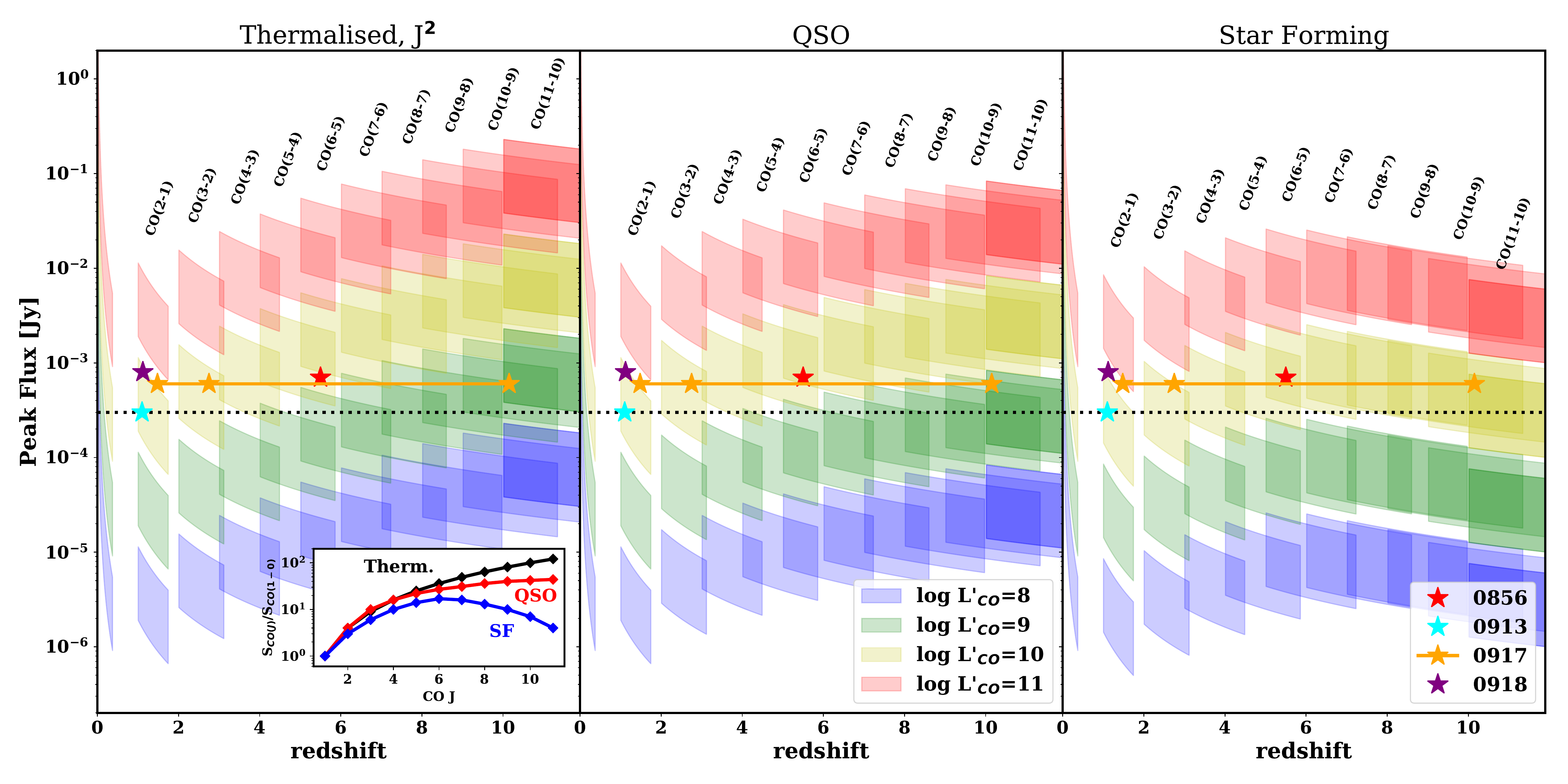}
    \caption{Prediction of CO peak flux  density for transitions within
    ALMA band 3 depending on (i) the CO SLED  (panels left to right), (ii) the intrinsic CO brightness 
     (middle insert) and the width of the line (shaded area), and (iii) the 
    on redshift  (the x-axis). The stars corresponding to our four sources per the lower right legend. For sources with more than 
    one possible redshift a line connects them.
    The CO SLEDs (presented in the inset in the lower left) follow a thermalized case (J$^2$, {\it left}),
    a typical quasar ({\it center}) and typical star forming
    galaxies ({\it right}), taken from \cite{carilli_cool_2013}.}
    \label{fig:CO_prediction}
\end{figure*}

With a measurement of the CO flux we can explore the properties of the molecular gas. In Fig.~\ref{fig:CO_prediction}, by assuming a spectral line energy distribution (SLED, the three panels, see insets) and an intrinsic \lpco, we can predict the integrated flux in each of the CO lines  as a function of redshift (shaded coloured areas). For each CO transition line, the horizontal span shows the frequency range of the given CO transition accessible with the ALMA band 3 (84-115\,GHz), and the vertical span, shows the line width range (limited here to 200-1000\,km\,s$^{-1}$). Note how the negative $k-$correction --- very similar to that observed at submm wavelength \citep[see ][]{blain_history_1999} --- allows for a relatively constant detection limit with increasing redshift.

We also report our  rms sensitivity ($\sigma$=0.3\,mJy/beam per 80\,MHz channel-width) as the black dotted line. This diagram directly shows that ALMA provides us with easy access down to an intrinsic CO luminosity \lpco=10$^{10}$\lpcounit, relatively independently of the redshift of the source. The final sensitivity is mainly driven by the shape of the SLED and therefore the excitation conditions of the source are the dominating factor, especially at the higher redshift end, the stronger the excitation properties, the lower in intrinsic luminosity we can reach. For a given source, the discrimination between the excitation mechanism will require observations of other CO transitions. One interesting point to note is independently of the intrinsic CO SLED, 0856  seems to lie around \lpco=10$^{10}$\lpcounit. For 0856 at $z=5.55$, we can estimate a \lpco=4*10$^{10}$\lpcounit. This value --- as well as the line width (930 km.s$^{-1}$) ---is very similar to other radio galaxies and quasars at lower and similar redshift \citep[][see their Fig. 5]{carilli_cool_2013}.

\section{Discussion}
\label{sec:disc}

\subsection{Selection Efficiency}

Out of four sources observed with ALMA, one can be considered as a bona fida detection of a powerful radio galaxy at $z=5.55$. For our definition of powerful radio galaxies ($L_{\rm 500\,MHz}>10^{27}\,$WHz$^{-1}$), with 0856 
we have already demonstrated the capability to break the 20-year old record from \cite{van_breugel_radio_1999}. 

Moreover, we have a very tentative candidate at $z=10.15$. 
We stress that this candidate still requires further observations to be securely confirmed. The two other sources are very likely lower redshift sources ($1<z<3$). While using some caution about low number statistics, we argue that we have detected at least 1/4 sources at $z>5$ from our full pilot sample and possibly up to 2/4 (assuming that the 0917 is at least at $z>5$) --- corresponding to a success rate of between $25-50\%$. 

This result is in comparison to \cite{saxena_search_2018}, who are probing lower flux densities at low frequency, and found 1 out of 32 sources at $z>5$ (the aforementioned $5.72$ radio galaxy). We note however they found four additional radio galaxies at $4<z<5$ \citep{saxena_nature_2019}. Our flux cut, F$_{\rm 151\,MHz}>0.4\,$Jy, \citep[in comparison to only using a upper limit of F$_{\rm 151\,MHz}\le 0.2\,$Jy such as the][]{saxena_discovery_2018}  ensures we select targets with relatively strong continuum emission. This admittedly prevents the detection of intrinsically lower luminosity objects but our high flux density limit is of prime importance for the future direct measure of HI absorption at such extreme redshifts.  Deeper surveys with the MWA will improve the signal-to-noise on our sources and make the model fitting across the MWA frequencies more secure.

\subsection{Comparison to other high-{\it z} AGN samples}

When comparing the NIR properties, the $K_s$ magnitude, is consistent with other samples of powerful radio galaxies (see Fig.~\ref{fig:Kz}), and therefore
we are preferentially selecting obscured, type 2 AGN.  This result is expected given the selection from weak/non-detection in $K-$band from the VIKING data. In comparison, the sample of bright QSOs from  \cite{banados_pan-starrs1_2016} is qualitatively brighter, in the $H=19-20$ mag (AB) range, as expected from their selection to be type 1, unobscured AGN. Note that this QSO sample does not follow the $K-z$ relation (Fig.~\ref{fig:Kz}) as the AGN outshines the host at optical/NIR wavelength  decreasing the observed $K-$band magnitude. Our selection is also biased to target massive systems: in the M$_*>10^{11-12}\,M_{\odot}$ range \citep{rocca-volmerange_radio_2004}. Interestingly, the \cite{saxena_nature_2019} sample tends to select similar or somewhat less massive systems (M$_*=10^{10.8-11.7}M_\odot$). Wider NIR photometric coverage is required to more accurately estimate stellar masses for our sample.

Our method is, by design, selecting the brightest radio sources in the sky (see Fig.~\ref{fig:radio_lum}),  unlike the recent sample described in \cite{saxena_nature_2019}. This makes our sample similar in nature to previously existing samples of powerful radio galaxies \citep[e.g. HeRG\'E,][]{seymour_massive_2007}.  Our two $z>5$ radio galaxies, GLEAM 0856 and GLEAM 0917, have observed-frame break frequencies around 1.4\,GHz. Hence we can compare the modeled spectral indices below this frequency to that of the Saxena sample. We find that our radio galaxies have less steep spectra than Saxena,
with $\alpha_{\rm low}\sim -0.9$ compared to $\alpha_{1.4GHz}^{151MHz}<-1.4$. However at higher frequencies 
our radio sources have $\alpha_{\rm high} \sim -1.6$ which meet the Saxena selection criteria. Hence our selection appears to favour radio galaxies with a higher rest-frame break-frequency than USS selected radio galaxies
and therefore potentially younger radio sources \citep[e.g.][]{turner_raise_2018}.

Interestingly, out of the hundreds of  optically bright AGN from the \cite{banados_pan-starrs1_2016} sample, a single one \citep[PSO J352.4034–15.3373 at $z=5.84\pm0.02$][]{banados_powerful_2018} is detected with a similar luminosity to GLEAM 0856.  This source has a comparable low frequency spectral index to our sample ($\alpha_{1.4GHz}^{150MHz}=-0.89$), but shows no evidence of steepening at higher redshift with a higher frequency spectral index of $\alpha_{3GHz}^{1.4GHz}=-0.78$. Potentially this up-turn could be indicative of a boosted radio emission from the core, expected for type 1 AGN. Or more likely it could simply be statistical uncertainty as the higher frequency spectral index is determined over a relatively narrow frequency range. 

 Our selection relies on low frequency spectral curvature which is only possible from the broad-band coverage of the MWA. This curvature is either due to synchrotron self-absorption processes or possibly free-free absorption from the host galaxy or circum-galactic medium. We note that PSO J352.4034–15.3373 does not show the low frequency curvature seen in our sample which could be due to its selection as an unobscured type 1 galaxy.

No systematic CO follow-up campaigns of samples of $z>5$ quasars exists as the low-J CO lines are redshifted outside of ALMA Band 3. We therefore compare qualitatively the 0856 properties to some smaller samples or individual objects. When compared to the hyper-luminous obscured quasars sample from \cite{fan_alma_2018}, 0856 appears relatively similar in term of CO brightness, but maybe on the higher side when compared to line-width. If compared with bright submm galaxies, such as the sample presented in \citep{weis_alma_2013} and comparing the CO brightness corrected from magnification, 0856 appears also similar. A comparison with the gas-rich star forming galaxies from \cite{tacconi_high_2010} indicates that 0856 CO brightness is in the higher-end of the sample. Finally, if compared to some sources of the HeRG\'E sample \citep{emonts_co1-0_2014,gullberg_alma_2016}, our derived CO brightnesses are a factor of few fainter, but this is mainly due to the sensitivity limit of the low-J CO surveys performed with ATCA.

\section{Conclusions}

We have presented a new and efficient method to identify and confirm powerful radio-loud AGN at $z>5$ by taking advantage of the new, low-frequency, all-sky GLEAM survey. We made use of the large frequency coverage (70-230\,MHz) to select sources from their spectral index and curvature. In our pilot project, we followed-up our best candidates
in the G09 field with VLT/HAWK-I, ATCA and ALMA. Out of our four sources, we successfully detected a source at $z=5.55$ and presented a very tentative candidate at $z=10.15$, which requires additional data for a robust confirmation\footnote{Further observations (to be presented in a forthcoming paper) are not yet able to robustly confirm the redshift of this source.}. Hence the efficiency of our method of finding $z>5.5$ radio galaxies is in the $25-50\%$ range (albeit from small number statistics).

From their radio luminosities, it appears clear that we preferentially select the most powerful radio sources, very similar to the already existing HeRG\'E sample \citep[L$_{\rm 500\,MHz}>10^{27}\,$W\,Hz$^{-1}$][]{seymour_massive_2007}. Their faint $K_s$ magnitude (preferably selecting  obscured type 2 AGN) place our object in the $K-z$ relation, known to pinpoint to the most massive system as any redshift, and consistent with M$_{\rm stel}$=10$^{11-12}M_{\odot}$ elliptical galaxies \citep{rocca-volmerange_radio_2004}. Moreover, as we suspect these galaxies harbour the most massive black holes, finding and confirming the redshifts of a larger sample of such sources 
could in the future present tight constraints on galaxy formation scenarios \citep[e.g.][]{haiman_constraints_2004,volonteri_rapid_2005}.

Our method shows that spectroscopy of bright molecular lines will be our only way to confirm routinely the redshift of these sources which are intrinsically fainter in $K_s$-band before the era of 30m-class telescopes and the {\it James Webb Space Telescope}. The hunt for powerful radio galaxies within the EoR continues.

\appendix
\section{GLEAM IDR3 vS the public data release}
\label{sec:IR3_vs_EGCv2}

Our initial source selection was based on the IDR3 version of the GLEAM extra-galactic catalogue available to the MWA consortium. If we apply our selection criteria (described in \S~\ref{sec:cand_selec}) to the public catalogue we obtain 71 sources (compared to 52) with 28 of each sample overlapping. The principle difference between the samples as a whole when we perform the MWA SED fitting is that the public data have on average an offset $\delta\beta=-0.3$ compared to IDR3. Hence, public sources will enter our selection criteria and some IDR3 sources will leave it. (this is the case for 0856 and 0917). Hence, the MWA sources are measured to be more curved in the public release than in IDR3. This can be explained by the lower frequency fluxes ($\nu<110\,$MHz) being fainter by up to 20\% in the public release (due to the improvements in data calibration and processing and a better characterisation of the related uncertainties). For this reason, we recommend using a lower $\beta$ interval to reproduce our selection with the public release (e.g. $-1.4<\beta<-0.6$). Our highest redshift candidates, 0856 and 0917 are recovered with these parameters. Note that the track for 8C\,1435+635 will not move as the curvature of this Northern hemisphere source is not defined by GLEAM data.

\begin{acknowledgements}

GD would like to thank Maria Strandet for her very useful and valuable input on the ALMA spectral extraction  and interpretation and Jess Broderick for careful reading of the manuscript. GD received the support of the ESO Scientific Visitor Programme. 

JA acknowledges financial support from the Science and Technology Foundation (FCT, Portugal) through research grants PTDC/FIS-AST/29245/2017, UID/FIS/04434/2013 and UID/FIS/04434/2019.

The National Radio Astronomy Observatory is the facility of the National Science Foundation operated under cooperative agreement by Associated Universities, Inc.

This paper makes use of the following ALMA data: ADS/JAO.ALMA\#2017.1.00719.S. ALMA is a partnership of ESO (representing its member states), NSF (USA) and NINS (Japan), together with NRC (Canada), MOST and ASIAA (Taiwan), and KASI (Republic of Korea), in cooperation with the Republic of Chile. The Joint ALMA Observatory is operated by ESO, AUI/NRAO and NAOJ. 

Based on observations made with ESO Telescopes at the La Silla Paranal Observatory under programme ID 0101.A-0571(A). GD would like to thank the user support department at ESO with the help provided with the HAWK-I pipeline. 

This scientific work makes use of the Murchison Radio-astronomy Observatory, operated by CSIRO. We acknowledge the Wajarri Yamatji people as the traditional owners of the Observatory site.

The Australia Telescope Compact Array is part of the Australia Telescope National Facility which is funded by the Australian Government for operation as a National Facility managed by CSIRO.

GAMA is a joint European-Australasian project based around a spectroscopic campaign using the Anglo-Australian Telescope. The GAMA input catalogue is based on data taken from the Sloan Digital Sky Survey and the UKIRT Infrared Deep Sky Survey. Complementary imaging of the GAMA regions is being obtained by a number of independent survey programmes including GALEX MIS, VST KiDS, VISTA VIKING, WISE, Herschel-ATLAS, GMRT and ASKAP providing UV to radio coverage. GAMA is funded by the STFC (UK), the ARC (Australia), the AAO, and the participating institutions. The GAMA website is \url{http://www.gama-survey.org/}. Based on observations made with ESO Telescopes at the La Silla Paranal Observatory under programme ID 179.A-2004.

This research made use of Astropy,\footnote{\tt http://www.astropy.org} a community-developed core Python package for Astronomy \citep{collaboration_astropy:_2013,collaboration_astropy_2018}. 

This research has made use of NASA Astrophysics Data System.

Facilities: ALMA, ATCA, JVLA, MWA, VISTA, VLT

Software: astropy, CASA, EAZY, MrMoose, numpy, scipy, sofia, sslf
\end{acknowledgements}

\bibliographystyle{pasa-mnras}
\bibliography{references}

\end{document}